\documentclass{jfm}

\usepackage{graphicx}
\usepackage{natbib}
\usepackage{epstopdf}
\usepackage{amsmath}
\usepackage{esdiff} 
\usepackage{subfig}
\usepackage{float}
\usepackage{relsize}
\usepackage{xcolor}
\usepackage{soul}

\ifCUPmtlplainloaded \else
  \checkfont{eurm10}
  \iffontfound
    \IfFileExists{upmath.sty}
      {\typeout{^^JFound AMS Euler Roman fonts on the system,
                   using the 'upmath' package.^^J}%
       \usepackage{upmath}}
      {\typeout{^^JFound AMS Euler Roman fonts on the system, but you
                   dont seem to have the}%
       \typeout{'upmath' package installed. JFM.cls can take advantage
                 of these fonts,^^Jif you use 'upmath' package.^^J}%
      }
  \else
  \fi
\fi

\ifCUPmtlplainloaded \else
  \checkfont{msam10}
  \iffontfound
    \IfFileExists{amssymb.sty}
      {\typeout{^^JFound AMS Symbol fonts on the system, using the
                'amssymb' package.^^J}%
       \usepackage{amssymb}%
         \let\leq=\leqslant
         
      }{}
  \fi
\fi

\ifCUPmtlplainloaded \else
  \IfFileExists{amsbsy.sty}
    {\typeout{^^JFound the 'amsbsy' package on the system, using it.^^J}%
     \usepackage{amsbsy}}
    {\providecommand\boldsymbol[1]{\mbox{\boldmath $##1$}}}
\fi

\newcommand\Real{\mbox{Re}} 
\newcommand\Imag{\mbox{Im}} 

\newsavebox{\astrutbox}
\sbox{\astrutbox}{\rule[-5pt]{0pt}{20pt}}

\DeclareMathOperator{\cotT}{\cot\left(\frac{\theta}{2}\right)}
\DeclareMathOperator{\thm}{\theta^{(m)}}
\DeclareMathOperator{\thj}{\theta^{(j)}}

\DeclareMathOperator{\cotTj}{\cot\left(\frac{\thj}{2}\right)}
\DeclareMathOperator{\cotTm}{\cot\left(\frac{\thm}{2}\right)}

\DeclareMathOperator{\bptwo}{\bar\Pi_2}
\DeclareMathOperator{\bpsix}{\bar\Pi_6}
\DeclareMathOperator{\bpsev}{\bar\Pi_7}

\DeclareMathOperator{\hatS}{\hat{S}_k^{(m)}}
\DeclareMathOperator{\hatC}{\hat{C}_k^{(m)}}

\DeclareMathOperator{\cost}{\cos\left(\beta^{(m)2}_n T_0\right)}
\DeclareMathOperator{\cosk}{\cos\left(\beta^{(m)2}_k T_0\right)}
\DeclareMathOperator{\coskin}{\cos({\beta^{(m)2}_k T_0})}
\DeclareMathOperator{\sint}{\sin\left(\beta^{(m)2}_n T_0\right)}
\DeclareMathOperator{\sink}{\sin\left(\beta^{(m)2}_k T_0\right)}
\DeclareMathOperator{\sinkin}{\sin({\beta^{(m)2}_k T_0})}

\DeclareMathOperator{\costz}{\cos\left(\beta^{(0)2}_n T_0\right)}
\DeclareMathOperator{\costo}{\cos\left(\beta^{(1)2}_n T_0\right)}
\DeclareMathOperator{\sintz}{\sin\left(\beta^{(0)2}_n T_0\right)}
\DeclareMathOperator{\sinto}{\sin\left(\beta^{(1)2}_n T_0\right)}

\title{Dynamics and instabilities of an arbitrarily clamped elastic sheet in potential flow with application to shape-morphing airfoils}
\shorttitle{Arbitrarily clamped elastic sheet in potential flow}
\author[Netanel Hassan, Shai B. Elbaz and Amir D. Gat]{N. Hassan, S.B. Elbaz and A.D. Gat}
\affiliation{Faculty of Mechanical Engineering, Technion - Israel Institute of Technology, Haifa 3200003, Israel}

\pubyear{2015} \volume{999} \pagerange{999--9999}
\date{\today}
\begin{document}
\maketitle

\abstract{
Shape-morphing airfoils have attracted much attention in recent years. They offer substantial drag reduction by comparison to conventional airfoils and a premise of superior aerodynamic performance. Such shape changing airfoils involve significant chord-wise elasticity, commonly akin to the aerodynamics of flags, sails as well as membrane wings and many natural flyers, but otherwise neglected in most conventional aircraft wing applications. In the current work, we model a shape-morphing airfoil as two, rear and front, Euler-Bernoulli beams connected to a rigid support at an arbitrary location along the chord. The setup is contained within a uniform potential flow field and the aerodynamic loads are modelled by thin airfoil theory. The aim of this work is to study the dynamics and stability of such soft shape-morphing configurations and specifically the modes of interaction between the front and rear airfoil segments. Initially we present several steady-state solutions, such as canceling of deflection due to aerodynamic forces and transition between two predefined cambers via continuous actuation of the airfoil. The steady results are validated by numerical calculations based on commercially available software. We then examine stability and transient dynamics by assuming small deflections and applying multiple-scale analysis to obtain a stability condition. The condition is attained via the compatibility equations of the  orthogonal spatial modes of the first-order correction. The results yield the maximal stable speed as a function of elastic damping, fluid density and location of clamping. The results show that the interaction between the front and rear segments is the dominant mechanism for instability for various discrete locations of clamping. Instabilities due to interaction dynamics between the front and rear segments become more significant as the location of clamping approaches the leading edge. Several transient dynamics are presented for stable and unstable configurations, as well as instability dynamics initiated by cyclic actuation at the natural frequency of the airfoil.

}

\section{Introduction}
Chord-wise elasticity of airfoils is common in biology, applications such as energy harvesting, sails, and more recently in the context of morphing wing sections which change their shape continuously \citep{macphee2016fluid,tiomkin2017stability,tang2018aeroelastic}. In this work we examine a chord-wise elastic airfoil in potential flow, clamped to a rigid supporting beam at an arbitrary location along the camber. The configuration can be viewed as a problem of a rear cantilevered elastic  sheet connected to an inverted front sheet. Under simplifying limits of thin-airfoil-theory and the Euler-Bernoulli beam model, we aim to study the dynamics and instabilities of such configurations. Of primary interest are aerodynamic interactions between the front- and rear-segments of the elastic airfoil, and the effect of such interactions on the onset of instability. 

The configuration examined in the current study is particularly relevant to shape-morphing airfoils involving chord-wise elasticity. Shape-morphing airfoils are currently extensively studied due to their potential to enhance performance of aircraft structures and energy harvesting systems \citep[see ][ among many others]{nguyen2015aeroelastic,takahashi2016development,moosavian2017parametric}. Common current approaches to shape-morphing of airfoils include piezoelectric actuation, shape memory alloys, pneumatic artificial muscles, as well as deployable and foldable structures (see detailed discussions in \cite{thill2008morphing}, \cite{barbarino2011review} and references therein). Realization of shape-morphing airfoils is commonly accompanied by increased chord-wise elasticity, which for sufficiently soft airfoils, may govern the aeroelastic response of the structure. 

In addition, chord-wise elasticity is a governing mechanism in tension-dominant   membrane wings, common in biology and  sail-like structures \citep{tiomkin2013membrane}. Previous research on membrane wings includes \cite{song2008aeromechanics}, who preformed an extensive experimental study, as well as comparison to a theoretical model of a wing camber under aerodynamic loading. Membrane wings were shown to provide greater lift, and improved lift slopes, due to modification of the camber at different angles of attack. \cite{alon2017steady} presented a framework for the analysis of the aeromechanics of membrane-wings.  A similar approach was used by \cite{tiomkin2017stability} to examine the stability of membrane wings as a function of the mass and tension of the membrane. Distributed actuation of membrane wings with variable compliance was studied experimentally by \cite{curet2014aerodynamic}, and later numerically by \cite{buoso2015electro}. Both works showed that increased aerodynamic efficiency may be obtained by leveraging  distributed actuation of membrane wings.

The current work is also relevant to the field of cantilevered elastic sheets, or flags, in uniform flow \citep{eloy2008aeroelastic,alben2008flapping,manela2009forced,alben2015flag,mougel2016synchronized}. The dynamics and stability of configurations in which the flow impinges over the clamped-end of the sheet have attracted significant interest \citep[we refer the reader to][ for a recent review]{shelley2011flapping}. In addition, in recent years interest emerged in the study of the inverted sheet configuration, where flow impinges over the free-end of the elastic sheet \citep{kim2013flapping,gilmanov2015numerical,gurugubelli2015self,sader2016stability,sader2016stabilitya}. These works are mainly motivated by the reduced flow speed required to induce self-oscillations in inverted sheets, which is relevant to energy harvesting applications.




The aim of the current work is to connect between regular and inverted cantilevered elastic sheets, and examine the effect of interaction between the sheets on the stability of the entire configuration.
The structure of this work is as follows: In \S 2 we define the problem and obtain the governing aeroelastic equation. In \S 3 we present several steady-state solutions based on regular asymptotic expansions and inverse solutions (solving the required actuation for pre-defined deformations). The results are compared with numerical calculations. In \S 4 we examine the effect of solid inertia on transient dynamics by applying multi-scale asymptotic expansions. We obtain stability requirements from the compatibility equations. Concluding remarks are provided in \S 5.

\section{Problem Formulation and Scaling}
\begin{figure}
\centering 
\includegraphics[width=1\textwidth]{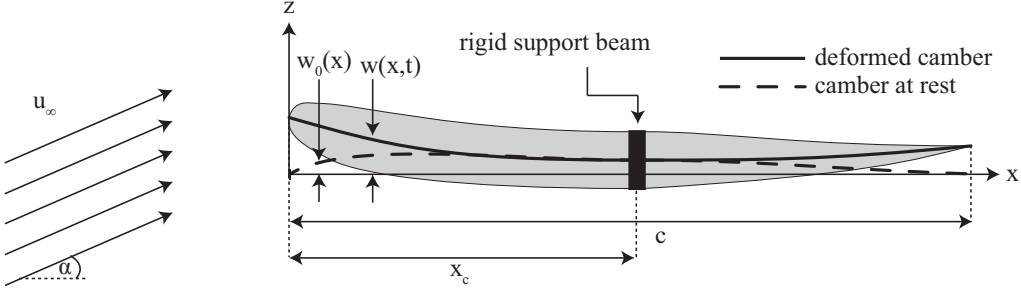}
\caption{Illustration of the examined configuration. $w_0$ (dashed line) is the camber of the airfoil at rest. $w$ (smooth line) is the total camber including elastic deformation. The black rectangle marks a rigid support beam to which the front and rear sections of the airfoil are clamped. The airfoil is sufficiently thin so as to allow the use of the thin airfoil and Euler-Bernoulli beam approximations.}
\label{Figure1}
\end{figure}

We examine the stability and dynamic response of an elastic two-dimensional airfoil actuated by the pressure-field of an external potential laminar flow. The airfoil elastic deformation is modelled by the Euler-Bernoulli equation, which is coupled to aerodynamic forces calculated by the thin airfoil theory. The examined configuration is illustrated in figure \ref{Figure1}.

We denote $w_0(x)$ as the camber at rest. $d_e(x,t)$ as the elastic deformation due to aerodynamic forces and $d_a(x,t)$ is forced actuation of the elastic airfoil. Total deformation from the initial state is $d(x,t)=d_e(x,t)+d_a(x,t)$ and total camber is $w(x,t)=w_0(x)+d(x,t)$. Chord-length is $c$. The $x$-coordinate is defined by the edges of camber at rest so that $w_0(0)=w_0(c)=0$. Angle-of-attack is $\alpha$ and velocity far from the airfoil is $(u_\infty\cos(\alpha),u_\infty\sin(\alpha))$. The airfoil is clamped by a rigid support beam at $x=x_c$. The parameters $s$ and $q$ denote sheet stiffness and aerodynamic loading per-unit-length in the perpendicular $z$-direction. The parameters $\mu_s$ and $r_d$ denote mass-per-area and elastic damping. 

The elastic deformation of the airfoil can be described by the Euler-Bernoulli equation
\begin{subequations} \label{governing_eq_dimenional}
\begin{equation}
\frac{\partial^2}{\partial x^2}\left[s\frac{ \partial^2 }{\partial x^2 } \left(w-w_0-d_a\right) \right]+r_d  \frac{\partial w}{\partial t}=-\mu_s \frac{ \partial^2 w}{\partial t^2 }+q 
\end{equation}
where $q$ is the aerodynamic load (per unit length in the $z$-direction). 

We aim to focus on the onset of instabilities, and simplify the governing equations by applying quasi-steady-state aerodynamic calculations. While such simplifications are commonly used \citep[e.g.][]{dowell1967generalized,fitt2001unsteady,mougel2016synchronized,sader2016large}, this assumption limits the validity of the results described in this work to configurations with negligible effects of vortex shedding (see discussion on the validity of this approximation at \S 2 of \cite{dowell1974aeroelasticity} and \S 5-6 of \cite{bisplinghoff2013aeroelasticity}). 
Under this approximation, the aerodynamic load $q$ can be presented by
\begin{multline}
\label{ThinWing} 
q=2\rho_\infty u_\infty^2 \Big\{\left[\alpha-\frac{1}{\pi}\int_0^\pi \left(\frac{\partial w}{\partial x} +\frac{1}{u_\infty} \frac{\partial w}{\partial t}\right) d\theta \right] \cotT+ \\ \sum_{n=1}^{\infty}\left[\frac{2}{\pi}\int_0^\pi \left(\frac{\partial w}{\partial x} +\frac{1}{u_\infty} \frac{\partial w}{\partial t} \right) \cos(\theta)d\theta\right] \sin(n\theta) \Big\}.
\end{multline}
\end{subequations}
where the auxiliary coordinate $\theta$ is defined by $x=c(1-\cos(\theta))/2$ \citep[e.g.][]{johnston2004review}.

Hereafter, Capital letters denote normalized variables and asterisk superscripts denote characteristic values (i.e., the normalized function $F$, is defined by $F=f/f^*$ where $f^*$ is a characteristic value of dimensional function $f$). We define the normalized axial coordinate $X={x}/{c}=(1-\cos(\theta))/{2}$, normalized time $T={t}/{t^*}$, normalized camber at rest $W_0(X)={w_0(x)}/{w_0^*}$, normalized actuation of the profile, $D_a(X,T)=d_a(x,t)/d_a^*$, normalized total deformation $D(X,T)= d(x,t)/{d^*}$, normalized rigidity $S(X)={s(x)}/{s^*}$, normalized damping for unit-length $B={r_d}/{r^*_d}$, and normalized mass-per-unit-length $ M_s(X)={\mu_s(x)}/{\mu_s ^*}$.

Substituting normalized variables and coordinates into (\ref{governing_eq_dimenional}) yields the normalized governing partial integro-differential equation 
\begin{multline}\label{main_equation}
\frac{\partial^2}{\partial X^2}\left[S(X)\frac{\partial^2 }{\partial X^2}\left(D-\Pi_1D_A\right)\right] +
\Pi_2 B \frac{\partial D}{\partial T}
+\Pi_3 M_s (X)\frac{\partial^2 D}{\partial T^2}\\
=\left[\Pi_4-\frac{1}{\pi} \int_0^\pi \left(\Pi_5 \frac{\partial W_0}{\partial X}+
\Pi_6 \frac{\partial D}{\partial X}+
\Pi_7 \frac{\partial D}{\partial T}  \right) d\theta\right]\cotT+\\
 \sum_{n=1}^{\infty}{\left[\frac{2}{\pi} \int_0^\pi \left(\Pi_5 \frac{\partial W_0}{\partial X}+
\Pi_6 \frac{\partial D}{\partial X}+
\Pi_7 \frac{\partial D}{\partial T}\right) \cos(n\theta)d\theta\right] \sin(n \theta)}
\end{multline}
and $\Pi_1-\Pi_7$ are dimensionless ratios defined by 
\begin{gather}
\Pi_1=\frac{d_a^*}{d^*},\quad \Pi_2=\frac{r^* c^4}{s^* t^*},\quad \Pi_3=\frac{\mu_s^* c^4 }{t^{*2} s^*},\quad \Pi_4=\frac{2\rho_\infty u_\infty^2  c^4\alpha}{s^* d^*},
\nonumber\\
\Pi_5=\Pi_4 \frac{w_0^*}{c\alpha},\quad \Pi_6=\Pi_4 \frac{d^*}{c\alpha},\quad \Pi_7=\Pi_4 \frac{d^*}{u_\infty t^*\alpha}.\label{ratios}
\end{gather}

The dimensionless number $\Pi_1$ represent the ratio of deformation due to actuation to the total deformation of the wing. Dimensionless numbers $\Pi_2-\Pi_7$ are all scaled by elastic bending forces, where $\Pi_2$ represents scaled damping, and $\Pi_3$ represents scaled inertia. $\Pi_4-\Pi_7$ represent scaled aerodynamic forces due to angle-of-attack of a flat camber $\Pi_4$, camber curvature at rest $\Pi_5$, camber deformation $\Pi_6$, and transient motion of the camber $\Pi_7$.

The governing integro-differential equation  (\ref{main_equation}) is supplemented by the boundary conditions at the support beam $X=X_c$ 
\begin{subequations}\label{BCs}
\begin{equation}\label{BC_XC}
D(X=X_c,T)=\frac{\partial D(X=X_c,T)}{\partial X}=0
\end{equation}
supplemented by zero moment and shear at $X=0$ and $X=1$
\begin{equation}\label{BC_ends}
\left[\frac{\partial^2 }{\partial X^2}(D-\Pi_1 D_A)\right]_{X=0,1}=\frac{\partial}{\partial X} \left[S\frac{\partial^2 }{\partial X^2}(D-\Pi_1 D_A)\right]_{X=0,1}=0
\end{equation}
\end{subequations}
and the initial conditions
\begin{equation}\label{IVs}
D(X,T=0)=F_1(X)\quad 
\frac{\partial D(X,T=0)}{\partial T}=F_2(X).
\end{equation}

\section{Steady-state solutions: Regular asymptotics, inverse solutions and numerical verification}
We start by examining the simple case of  steady-state deformations, and compare the results to numerical calculations by commercially available code (COMSOL Multiphysics \textregistered 5.2a). In this limit, the characteristic time-scale $t^*$ is required to be sufficiently large so that
\begin{equation}
    \Pi_2,\, \Pi_3,\, \Pi_7 \ll 1,
\end{equation}
reducing (\ref{main_equation})  to the quasi-steady equation
\begin{multline}\label{SteadyGov}
\frac{\partial^2}{\partial X^2}\left[S(X)\frac{\partial^2 }{\partial X^2}\left(D-\Pi_1D_A\right)\right] 
=\left[\Pi_4-\frac{1}{\pi} \int_0^\pi \left(\Pi_5 \frac{\partial W_0}{\partial X}+
\Pi_6 \frac{\partial D}{\partial X}
 \right) d\theta\right]\cotT+\\
 \sum_{n=1}^{\infty}{\left[\frac{2}{\pi} \int_0^\pi \left(\Pi_5 \frac{\partial W_0}{\partial X}+
\Pi_6 \frac{\partial D}{\partial X}
\right) \cos(n\theta)d\theta\right] \sin(n \theta)},
\end{multline}
along with initial and boundary conditions (\ref{BCs}).
(The requirements for stability of such steady-state solutions will be examined in \S 4.)

\subsection{Asymptotic expansion for the limit of small deflections}
We define the small parameter
\begin{equation}
    \varepsilon=\frac{\Pi_6}{\Pi_4+\Pi_5}
\end{equation}
(where $\Pi_2,\, \Pi_3,\, \Pi_7 \ll \varepsilon \ll 1$) representing the ratio between aerodynamic forces due to elastic deflection of the camber and the sum of aerodynamic forces due to the angle-of-attack ($\Pi_4$) and the undeformed camber ($\Pi_5$).

Substituting the expansion 
\begin{equation}
D = \sum_{n=0}^{\infty} \varepsilon^n D_n
\label{asymptotic_expansion}
\end{equation}
into (\ref{SteadyGov}), along with $\Pi_6=\varepsilon (\Pi_4+\Pi_5)$,  yields the leading-order equation governing $D_0$
\begin{subequations}\label{Steady_Asymp}
\begin{multline}\label{le_ord}
\frac{\partial^2}{\partial X^2}\left[S(X)\frac{\partial^2 }{\partial X^2}\left(D_0-\Pi_1D_A\right)\right] 
=
\left[
\Pi_4-\frac{1}{\pi}
\int_0^\pi \left(
\Pi_5 \frac{\partial W_0}{\partial X}
\right) d\theta
\right]\cotT \\+
 \sum_{n=1}^{\infty}
 {\left[\frac{2}{\pi} \int_0^\pi
 \left(
 \Pi_5 \frac{\partial W_0}{\partial X}
\right)
\cos(n\theta)d\theta\right] \sin(n \theta)},
\end{multline}
and higher orders $O(\varepsilon^n)$ equations governing $D_n$ are given by 
\begin{multline}\label{n_ord}
\frac{\partial^2}{\partial X^2}\left(S\frac{\partial^2 D_n}{\partial X^2}\right)=
(\Pi_4+\Pi_5) \Big[
\left(-\frac{1}{\pi} \int_0^\pi \frac{\partial D_{n-1} }{\partial X}
  d\theta\right)\cotT+ \\
 \sum_{l=1}^{\infty}{\left(\frac{2}{\pi} \int_0^\pi  \frac{\partial D_{n-1} }{\partial X}
 \cos(l\theta)d\theta\right) \sin(l \theta)}
\Big].
\end{multline}
\end{subequations}
Since the integrals in (\ref{Steady_Asymp}) depend on known terms, $D_n$ may be computed from (\ref{Steady_Asymp}) by direct integration with regards to $X$. Due to the clamping at $X=X_c$, the curvature will be discontinuous. Applying boundary conditions at the free-ends $X=0,1$,
the leading-order curvature due to elastic deflections is
\begin{subequations}\label{Steady_Asymp_curve}
\begin{multline}\label{le_sol_pertubations}
\frac{\partial^2 D_0}{\partial X^2}=\Pi_1\frac{\partial^2 D_A}{\partial X^2}+\frac{1}{S}\int_0^\theta 
\int_0^{\theta^{*}}
\Bigg[
\left(
\frac{\Pi_4}{4}-\frac{\Pi_5}{4\pi} 
\int_0^\pi {\frac{\partial W_0}{\partial X} }d\theta
\right) \cot \left(\frac{{\theta^{**}}}{2}\right)+
\\+ \frac{\Pi_5}{2\pi }
\sum_{n=1}^\infty
\left(
\int_0^\pi {\frac{\partial W_0}{\partial X}\cos(n\theta) }d\theta
\right)
\sin(n{\theta^{**}})
\Bigg]
\sin({\theta^{**}})d{\theta^{**}}
\sin({\theta^{*}})d{\theta^{*}}
\\
\frac{\pi H(X-X_c)}{4S} \left\{
\Pi_4 \left( \cos(\theta) -\frac{1}{2} \right)
+\frac{\Pi_5}{\pi} \int_0^\pi
\frac{\partial W_0}{\partial X}
\left[\left( \cos({\theta^{*}}) -1 \right) \cos(\theta)-
\frac{\cos(2{\theta^{*}})-1}{2}\right]d{\theta^{*}}
\right\},
\end{multline}
and higher-order corrections are given by
\begin{multline}\label{he_sol_pertubations}
\frac{\partial^2 D_n}{\partial X^2}=
\frac{\Pi_4+\Pi_5}{4\pi S}
\int_0^\theta 
\int_0^{\theta^{*}}
\Bigg[
\left(
- \int_0^\pi {\frac{\partial D_{n-1}}{\partial X} }d\theta
\right) \cot \left(\frac{{\theta^{**}}}{2}\right)
\\+2 \sum_{l=1}^\infty
\left(
\int_0^\pi {\frac{\partial D_{n-1}}{\partial X}\cos(l\theta) }d\theta
\right)
\sin(l{\theta^{**}})
\Bigg]
\sin({\theta^{**}})d{\theta^{**}}
\sin({\theta^{*}})d{\theta^{*}}
+\\ 
\frac{\Pi_4+\Pi_5}{4S}
\left\{\int_0^\pi
\frac{\partial D_{n-1}}{\partial X}
\left[\left( \cos({\theta^{*}}) -1\right) \cos(\theta)-
\frac{\cos(2{\theta^{*}})-1}{2}\right]d{\theta^{*}}
\right\}
H(X-X_c),
\end{multline}
where $H(X)$ is the Heaviside step function and $\theta^{*}$, $\theta^{**}$ are auxiliary integration coordinates.
\end{subequations}

Results (\ref{Steady_Asymp_curve}) for the steady deflection of an actuated airfoil are presented in figure \ref{numeric_validations}a, and compared to numerical simulations of potential flow over an elastic NACA-2412 airfoil. Good agreement is evident. Details regarding the physical and geometric parameters,  the numerical scheme, as well as discussion, are presented  below in \S 3.3.

\subsection{Inverse solutions}
For known fluid and airfoil properties, distributed  actuation can be applied to achieve reduced aeroelastic deflection of soft airfoils, or to create transition between two predefined cambers. By setting the deformation $D$ and solving for the actuation $D_A$, it is possible to solve (\ref{SteadyGov}) without application of asymptotic expansions.

Cancellation of steady aeroelastic deflection by distributed actuation is immediately calculated from (\ref{main_equation}) by setting $D=0$, yielding  $D_A$ by
\begin{multline}\label{trans_sol}
 \frac{\partial^2 D_A}{\partial X^2}=
-\frac{1}{\Pi_1 S(X)}\int_0^\theta 
\int_0^{\theta^{*}}
\Bigg[
\left(
\frac{\Pi_4}{4}-\frac{\Pi_5}{4\pi} 
\int_0^\pi {\frac{\partial W_0}{\partial X} }d\theta
\right) \cot \left(\frac{{\theta^{**}}}{2}\right)
\\+ \frac{\Pi_5}{2\pi}
\sum_{n=1}^\infty
\left(
\int_0^\pi {\frac{\partial W_0}{\partial X}\cos(n\theta) }d\theta
\right)
\sin(n{\theta^{**}})
\Bigg]
\sin({\theta^{**}})d{\theta^{**}}
\sin({\theta^{*}})d{\theta^{*}}
+\frac{H(X-X_c)}{4\Pi_1 S(X)}\cdot\\ \left\{
\Pi_4 \left( \cos(\theta) -\frac{1}{2} \right)
+\Pi_5 \int_0^\pi
\frac{\partial W_0}{\partial X}
\left[\left( \cos({\theta^{*}}) -1\right) \cos(\theta)-
\frac{\cos(2{\theta^{*}})-1}{2}\right]d{\theta^{*}}
\right\}.
\end{multline}
(Additional integration coefficients are obtained from clamping boundary conditions (\ref{BC_XC}) at $X=X_c$.)

Similarly, transition between two predefined cambers $W_1(X)$ and $W_2(X)$ can be readily obtained by the following scheme: (I) setting $D_A=0$ and $W_0=W_1-D\Pi_6/\Pi_5$ in (\ref{SteadyGov}) and solving for $D$. In this case the RHS integral in (\ref{SteadyGov}) contains only $W_1$. Hence, the equation is no longer an integro-differential equation and $D$ can be calculated by integration. From $D$ the initial camber at rest $W_0$ is obtained for which the profile $W_1$ is acheived for $D_A=0$.  (II) After calculating $W_0$, $D_A$ can be calculated by setting $D\Pi_6/\Pi_5=W_2-W_0$ into (\ref{SteadyGov}) and solving for $D_A$ yielding camber $W_2$.

Figure \ref{numeric_validations}b presents example of cancellation of aeroelastic deformation, and figure \ref{numeric_validations}c presents  transition between two predefined cambers. Both configurations are compared with numerical calculations, showing good agreement. A detailed description of the numerical scheme, physical and geometric parameters, and discussion, are presented below.

\subsection{Numerical validation}
In this section we present illustrative examples of results from \S 3.1 and \S 3.2, and compare the analysis to numerical calculations. We focus on a NACA-2412 airfoil geometry with chord of $c=1m$, clamped at $x_c=0.25m$  Young's modulus $E=8MPa$. Solid density is $\rho_s=1600 kg/m^3$ and the density of the fluid is $\rho_\infty=1.006 kg/m^3$ (taken from standard atmosphere model for altitude of $2 km$). The angle-of-attack is $\alpha=5^\circ$, and the uniform potential flow velocity is $u_\infty=40 m/s$.

The specific method in which actuation is achieved is not essential to the analysis, and here we arbitrarily chose camber actuation by a distribution of pressurized internal chambers, commonly used in the field of soft robotics. Such actuation approach is known as embedded-fluidic-networks or pneumatic-artificial-muscles, \citep{thill2008morphing}. A description of the relation between the function $d_a$ and the pressure and geometry of the chambers, is presented in   \cite{matia2015dynamics} as a long-wave approximation by
\begin{equation}
    \frac{\partial^2d_a}{\partial x^2}=\phi(x)\psi(p_c,x),
\end{equation}
where $\phi$ represents channel density ($1/\phi(x)$ is the distance between the channels) and $\psi(p_c,x)$ is the total change in slope $\partial d_a/\partial x$ due to the actuated channel, where $p_c$ is the pressure within the channels. In the presented calculations the channel cross-section is a circle of diameter of $h/5$ with center located $2h/7$ above the midplane, where $h=h(x)$ is the local thickness of the airfoil. For the above parameters, and the limit of $p_c/E\ll1$, $\psi$ is approximated as $\psi\approx 0.1741(p_c/E)
$ \citep[see][for detailed description] {matia2015dynamics}.

The numerical calculations utilized commercially available code (COMSOL Multiphysics \textregistered 5.2a), with grid consisting of $10^3$ first-order unstructured triangular elements with average element quality of $0.94$ for the fluidic domain and $10^3$ second-order unstructured triangular elements with average element quality of $0.9$ for the solid domain. The size of the rectangular domain was $8c\times10c$, rotated by $\alpha$ so that the velocity condition at the front boundary is perpendicular to the boundary.
The model included $10^4$ degrees of freedom. All our solutions converged by at least $6$ orders of magnitude from the value given at the initial condition. In the first step the solver created the flow field, allowing deformations to be created and become stable. Then, in the second step, internal pressure was applied within the chambers.

\begin{figure}
\centering 
    \includegraphics[width=1\linewidth]{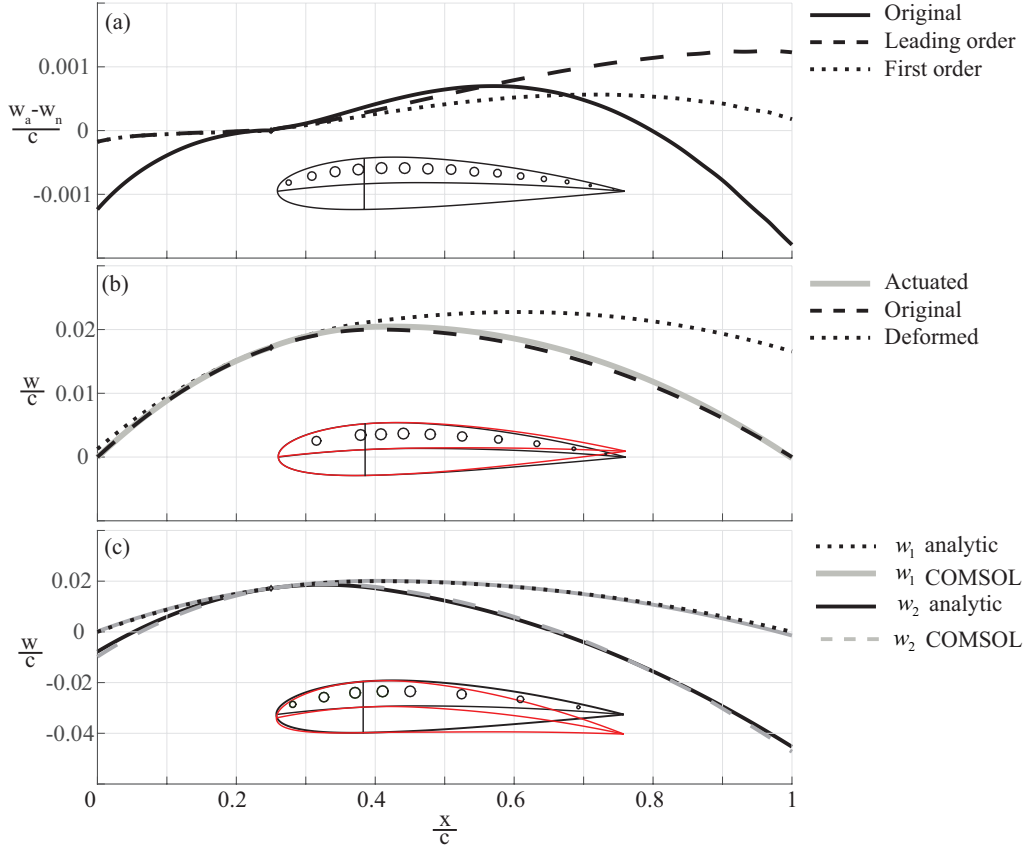}
    \caption{Numeric validations for the results in \S 3.1 and \S 3.2. Small inserts demonstrate airfoils' shapes, with embedded-fluidic-network distributed actuation (circles), as presented in \S 3.3.
    Panel (a) presents results for the asymptotic expansion in \S 3.1 in the form of spatial error of the leading and the first order compared with the numerical solution ($w_a$ is the analytic solution, $w_n$ is numerical calculation).
    Panel (b) presents results for cancellation of aeroelastic deformation using actuation as the cambers shapes of the original, the deformed and the final shaped, where actuation is applied. Good agreement can be seen.
    Panel (c) presents results of using actuation to transform NACA2412 airfoil into NACA4412 airfoil. Good agreements are seen between cambers shapes of the analytic and numerical solutions.}\label{numeric_validations}
\end{figure}

Figure \ref{numeric_validations} presents comparison between analytic and numerical results. Panel (a) compares the difference between the numerical calculation $w_n$ and the analytic results $w_a$ computed by the asymptotic scheme (\ref{Steady_Asymp_curve}) for distributed actuation of $\partial^2 d_a/\partial x^2=1.9425$. The channel distribution is presented at the insert, where the the channels are pressurized at $p_c=150kPa$. Difference between the results is presented for no-correction (smooth line), leading-order correction (dashed line) and first-order correction (dotted line). The asymptotic scheme clearly reduces the discrepancy between the analysis and the numerical computation with increasing order of the correction terms.

Panels (b) and (c) in figure \ref{numeric_validations} present comparison between numerical calculations and analytic results by  inverse calculations, described in \S 3.2. Panel (b) presents deformation cancellation, which determines the required actuation, and thus the geometry of the chambers, (see insert in panel b; channels are pressurized at $p_c=146.5kPa$). The camber at rest is marked by a smooth line, the deformed unactuated camber is denoted by a dashed line. The numerical calculation for a camber actuated according to (\ref{trans_sol}) is marked by a dotted line. Clear agreement between the numerical calculation and the analytic results is evident. Panel (c) presents comparisons for transformation of NACA2412 camber to NACA4412 camber (channels are pressurized at $p_c=896.2kPa$). Good agreement between the analysis and numerical computations is presented for this case as well. 



\section{Transient solutions: Multi-scale expansion and stability analysis}
This section examines transient dynamics and stability at the limit of small structural damping, defined as
\begin{equation}
\Pi_2=\varepsilon\bptwo,
\end{equation}
and small aerodynamic forces due to elastic deflection, defined as  
\begin{equation}
\Pi_6=\varepsilon \bpsix ,\quad \Pi_7=\varepsilon \bpsev
 \end{equation}
where $\bar \Pi_2, \bar \Pi_6, \bar \Pi_7\sim O(1)$. For this limit order-of-magnitude yields
\begin{equation}
t^*=c^2\sqrt{\frac{\mu_s}{s^*}},\quad d^*=\frac{2\rho_\infty U_\infty^2*c^3}{s^*}\max{[c\alpha,w_0^*]}
 \end{equation}
and thus hereafter  $\Pi_3=1$. In addition, for simplicity, we focus on constant sheet stiffness and mass-per-unit-length: $S(X)=1,\,M_s(X)=1$. 

Dynamics and stability requirements will be obtained via the compatibility equations of the multi-scale expansions for the different spatial oscillation modes.  We refer the reader to \cite{hinch1991perturbation} and \cite{bender2013advanced} for discussion of the method of multi-scale asymptotic expansions.

\subsection{Multiple-scale asymptotic expansion}
We introduce a slow time-scale $T_1=\varepsilon T$ as well as an asymptotic expansion with regards to $\varepsilon$,
\begin{subequations}\label{expansion}
\begin{equation}
D(X,T)=D_0(X,T_0,T_1)+\varepsilon D_1(X,T_0,T_1)+O(\varepsilon^2) 
\end{equation} 
and thus
\begin{equation}
\frac{\partial}{\partial T}=\frac{\partial}{\partial T_0}+\varepsilon \frac{\partial}{\partial T_1}. 
\end{equation}
\end{subequations}

Substituting (\ref{expansion}) into (\ref{main_equation}), the leading- and first-order equations are
\begin{multline}\label{Leading_order_equation}
\frac{\partial^4 D_0}{\partial X^4}+
 \frac{ \partial^2 D_0}{\partial T_0^2}
=\Pi_1 \frac{\partial^4 D_A}{\partial X^4}
+\Pi_4 \cotT +
\\\Pi_5 \left[
\left(
-\frac{1}{\pi}\int_0^\pi \frac{\partial W_0}{\partial X}d\theta
\right)
\cotT
+\sum_{n=1}^\infty
\left(
\frac{2}{\pi}\int_0^\pi \frac{\partial W_0}{\partial X} \cos(n\theta) d \theta
\right) \sin(n\theta)
\right]
\end{multline}
and
\begin{multline}\label{first_order_equation}
\frac{\partial^4 D_1}{\partial X^4}+
 \frac{ \partial^2 D_1}{\partial T_0^2}
=-\bptwo B \frac{\partial D_0}{\partial T_0}
-2 \frac{\partial^2 D_0}{\partial T_0 \partial T_1}\\
+\bpsix
\left[
\left(
-\frac{1}{\pi}\int_0^\pi \frac{\partial D_0}{\partial X}d\theta
\right)
\cotT
+\sum_{n=1}^\infty
\left(
\frac{2}{\pi}\int_0^\pi \frac{\partial D_0}{\partial X} \cos(n\theta) d \theta
\right) \sin(n\theta)
\right]\\
+\bpsev
\left[
\left(
-\frac{1}{\pi}\int_0^\pi \frac{\partial D_0}{\partial T_0}d\theta
\right)\cotT
+\sum_{n=1}^\infty
\left(
\frac{2}{\pi}\int_0^\pi \frac{\partial D_0}{\partial T_0} \cos(n\theta) d \theta
\right) \sin(n\theta)
\right].
\end{multline}

\begin{subequations}
Leading-order boundary and initial conditions for $D_0$ are identical to (\ref{BCs}). The first-order equation is supplemented by the homogeneous first-order boundary conditions
\begin{equation}
    D_1\biggr\rvert_{X=X_C}=\frac{\partial D_1}{\partial X}\biggr\rvert_{X=X_C} = \frac{\partial^2 D_1}{\partial X^2}\biggr\rvert_{X=0,1}=
    \frac{\partial^3 D_1}{\partial X^3}\biggr\rvert_{X=0,1}=0,
\end{equation}
and the initial condition
\begin{equation}
D_1(X,T_0=0,T_1=0)=0\qquad 
\frac{\partial D_1(X,T_0=0,T_1=0)}{\partial T_0}=0.
\end{equation}
\end{subequations}

The airfoil is connected to a rigid support at $X=X_c$ and we denote hereafter the upstream segment $0<X<X_c$ by the superscript $(0)$ and the downstream segment $X_c<X<1$ by the superscript $(1)$. In addition we define the auxiliary coordinates  $\xi^{(m)}$  as
\begin{equation}\label{xi_def}
\xi^{(m)}=\frac{X-X_c}{m-X_c}=
\frac{\cos(\theta_c)-\cos(\theta)}{2m-1+\cos(\theta_c)},
\end{equation}
where $m\in\{0,1\}$.

\subsection{Leading-order solution}
While rather tedious, the solution of the leading-order equation can be obtained by standard methods of homogenization and separation of variables \cite{kreyszig2010advanced}. Thus, the results are directly presented.

The leading-order solution $D_0^{(m)}$ can be expressed by
\begin{equation}\label{LE_sol}
 D_0^{(m)}=\sum_{n=1}^\infty  \Theta_n^{(m)}(T_0,T_1) \Xi_n(\xi^{(m)})+D_{0,SS}(\xi^{(m)})+V(\xi^{(m)},T_0,T_1).
\end{equation}
The function $V^{(m)}$ is the boundary condition homogenization function, given by \begin{equation}\label{V_def}
V^{(m)}=-\frac{\Pi_1 (m-X_c)^2\xi^{(m)2}}{6}
\left[
3\frac{\partial ^2 D_A}{\partial X^2}\biggr\rvert_{X=m}
-(3-\xi^{(m)})\left[m+(-)^m X_c\right]\frac{\partial ^3 D_A}{\partial X^3}\biggr\rvert_{X=m}
\right].
\end{equation}
The $\Xi_n(\xi^{(m)})$ functions are the spatial eigenmodes 
\begin{equation}
\Xi_n(\xi^{(m)})=\cos(\lambda_n \xi^{(m)})-\cosh(\lambda_n\xi^{(m)})-\frac{\cos(\lambda_n)+\cosh(\lambda_n)}{\sin(\lambda_n)+\sinh(\lambda_n)} \left[\sin(\lambda_n\xi^{(m)})-\sinh(\lambda_n\xi^{(m)})\right]
\end{equation}
and $\Theta_n^{(m)}(T_0,T_1)$ are temporal eigenmodes
\begin{subequations} \label{Theta_def}
\begin{equation} \label{Theta_hom}
\Theta_n^{(m)}(T_0,T_1)=C_n^{(m)}(T_1)\cost
+S_n^{(m)}(T_1)\sint + \Theta_{n,A}^{(m)},
\end{equation}
where $\Theta_{n,A}^{(m)}$ represents the influence of the actuation, reads
\begin{equation}\label{Theta_a}
\Theta_{n,A}^{(m)}=
\frac{1}{\beta^{(m)2}_n}\sint
\circledast
\left[\int_0^1
\left(\Pi_1 \frac{\partial^4 D_A}{\partial X^4}
- \frac{\partial^2 V^{(m)}}{\partial T_0^2}
\right) \Xi_n(\xi^{(m)})d\xi^{(m)}
\right]
\end{equation}
($\circledast$ denotes convolution with regards to $T_0$).
\end{subequations}

\begin{subequations}\label{IVS_C&S}
Substituting initial conditions (\ref{IVs}c) into $D_0^{(m)}$, we obtain the initial values of $C_n^{(m)}(T_1)$, $S_n^{(m)}(T_1)$ by
\begin{equation}
    C_n^{(m)}(0)=\int_0^1 \left[F_1(\xi^{(m)})-D_{0,SS}(\xi^{(m)})-V(\xi^{(m)},0,0)\right] \Xi_n(\xi^{(m)}) d\xi^{(m)}
\end{equation}
\begin{equation}
    S_n^{(m)}(0)=\int_0^1 \left[F_2(\xi^{(m)})- \frac{\partial V(\xi^{(m)},0,0)}{\partial T_0}\right] \Xi_n(\xi^{(m)}) d\xi^{(m)}
\end{equation} 
\end{subequations}

The eigenvalues $\lambda_n$ are the positive roots of the transcendental equation \begin{equation}\label{trans_eq}
\cos(\lambda_n)\cosh(\lambda_n)=-1
\end{equation}
and  $\beta_n^{(m)}$ are related to the solution of the transcendental equation (\ref{trans_eq}) via
\begin{equation} \label{freq_relation}
\beta_n^{(m)}= \frac{\lambda_n}{m+(-)^m X_c}.
\end{equation}
Finally, the function $D_{0,SS}$ is the steady-state leading-order solution, which is readily obtained from
\begin{subequations}
\begin{multline}
 \frac{\partial^4 D_{0,SS}}{\partial X^4}=
\Pi_4 \cotT +\\ \Pi_5 \left[
\left(
-\frac{1}{\pi}\int_0^\pi \frac{\partial W_0}{\partial X}d\theta
\right)
\cotT
+\sum_{n=1}^\infty
\left(
\frac{2}{\pi}\int_0^\pi \frac{\partial W_0}{\partial X} \cos(n\theta) d \theta
\right) \sin(n\theta)
\right]   
\end{multline}
with boundary conditions
\begin{equation}
\frac{\partial^2 D_{0,SS}}{\partial X^2}\biggr\rvert_{X=m}=\frac{\partial^3 D_{0,SS}}{\partial X^3}\biggr\rvert_{X=m}=
D_{0,SS}\biggr\rvert_{X=X_c}=\frac{\partial D_{0,SS}}{\partial X}\biggr\rvert_{X=X_c}=0.
\end{equation}
\end{subequations}

\subsection{Identification of secular-terms} 
Substituting the above leading-order solution into the first-order correction, equation (\ref{first_order_equation}) becomes
\begin{multline} \label{first_order}
\frac{\partial^4 D_1^{(m)}}{\partial X^4}
+ \frac{\partial^2 D_1^{(m)} }{\partial T_0^2}=\\
-2 \sum_{n=1}^\infty \frac{\partial^2 \Theta_n^{(m)} }{\partial T_0 \partial T_1} \Xi_n(\xi^{(m)})
-\bptwo B \frac{\partial V^{(m)}}{\partial T_0} -\bptwo \sum_{n=1}^\infty B_n \frac{\Theta_n^{(m)}}{\partial T_0}\Xi_n(\xi^{(m)})
\\+
\bpsix \Biggr[-\frac{1}{\pi}
\left[
\sum_{n=1}^\infty
\left(\Theta_n^{(0)} \int_0^{\theta_c}\frac{\Xi_n^{(0)}}{\partial X}d\theta+
\Theta_n^{(1)} 
\int_{\theta_c}^\pi \frac{\Xi_n^{(1)}}{\partial X}d\theta
\right)
\right]\cotTm 
\\+\frac{2}{\pi} \sum_{l=1}^\infty
\left[ \sum_{n=1}^\infty
\left(\Theta_n^{(0)}
\int_0^{\theta_c}\frac{\Xi_n^{(0)}}{\partial X} \cos(l\theta) d\theta
+\Theta_n^{(1)}
\int_{\theta_c}^\pi\frac{\Xi_n^{(1)}}{\partial X} \cos(l\theta) d\theta
\right)
\right]
\sin(l\thm)
\Biggr]\\
+\bpsev \Biggr[-\frac{1}{\pi}
\left[
\sum_{n=1}^\infty
\left(\frac{\partial \Theta_n^{(0)}}{\partial T_0} 
\int_0^{\theta_c}\Xi_n^{(0)}d\theta
+\frac{\partial \Theta_n^{(1)}}{\partial T_0}
\int_{\theta_c}^\pi \Xi_n^{(1)}d\theta
\right)
\right]\cotTm
\\+\frac{2}{\pi} \sum_{l=1}^\infty
\left[ \sum_{n=1}^\infty
\left(\frac{\partial \Theta_n^{(0)}}{\partial T_0} 
\int_0^{\theta_c}\Xi_n^{(0)}\cos(l\theta) d\theta
+\frac{\partial \Theta_n^{(1)}}{\partial T_0}
\int_{\theta_c}^\pi \Xi_n^{(1)}\cos(l\theta) d\theta
\right)
\right]
\sin(l\thm)
\Biggr]\\
-\frac{
\cotTm}{\pi}\int_0^\pi
\left[
\bpsix \left(\frac{\partial D_{0,SS}}{\partial X}+\frac{\partial V}{\partial X}
\right)
+\bpsev \frac{\partial V}{\partial T_0}
\right]
d\theta
\\ + 
\sum_{l=1}^\infty 
\sin(l\thm)\frac{2}{\pi} \int_0^\pi
\left[
\bpsix \left(\frac{\partial D_{0,SS}}{\partial X}+\frac{\partial V}{\partial X}
\right)
+\bpsev \frac{\partial V}{\partial T_0}
\right]
\cos(l\theta) d\theta.
\end{multline}

Since the boundary conditions for $D_1^{(m)}$ are homogeneous, we suggest solution of the form 
\begin{equation}
D_1^{(m)}(\xi^{(m)},T_0,T_1)=\sum_{n=1}^\infty\Xi_n(\xi^{(m)})Q_n^{(m)}(T_0,T_1)
\end{equation}
where $\Xi_n(\xi)$ are identical to the eigenmodes of the leading-order solution. Hence, the LHS of equation (\ref{first_order}) becomes
\begin{equation}
\frac{\partial^4 D_1^{(m)}}{\partial X^4}
+ \frac{\partial^2 D_1^{(m)} }{\partial T_0^2}=
\sum_{n=1}^\infty \left[
Q_n^{(m)}\beta^{(m)4}_n+\frac{\partial^2 Q_n^{(m)}}{\partial T_0^2}
\right] \Xi_n(\xi^{(m)}).
\end{equation}

We define the operator 
\begin{multline}\label{operator_defenition}
    \mathcal{N}\left[{F(\xi^{(j)})};{G(\xi^{(i)})}\right]=
    \int_0^1\bigg[-\frac{1}{\pi}
    \left(
    \int_{\theta^{(j)}(\xi^{(j)}=1-j)}^{\theta^{(j)}(\xi^{(j)}=j)}
    F(\xi^{(j)}) d\theta^{(j)}
    \right)\cotTj \\
    +\frac{2}{\pi}\sum_{k=1}^\infty
    \left(
    \int_{\theta^{(j)}(\xi^{(j)}=1-j)}^{\theta^{(j)}(\xi^{(j)}=j)}
    F(\xi^{(j)}) \cos(k\theta^{(j)})d\theta^{(j)}
    \right)\sin(k\theta^{(i)})
    \bigg]G(\xi^{(i)}) d\xi^{(i)}
\end{multline}
and for brevity, we hereafter denote the auxiliary scalars $\mathcal{I}^{k,(i)}_{n,(j)}$ and $\mathcal{J}^{k,(i)}_{n,(j)}$ as \begin{subequations}\label{Aux_scalars}
\begin{equation}
\mathcal{I}^{k,(i)}_{n,(j)}=
\mathcal{N}\left[{\Xi_n(\xi^{(j)})};{\Xi_k(\xi^{(i)})}\right] 
\end{equation}
and
\begin{equation}
\mathcal{J}^{k,(i)}_{n,(j)}=
\mathcal{N}\left[\frac{\partial \Xi_n(\xi^{(j)})}{\partial X};\Xi_k(\xi^{(i)})\right]
\end{equation}
\end{subequations}
representing interaction between the sheets' structural modes and the accompanying pressure distribution due to quasi-static aerodynamic modes.

Multiplying equation (\ref{first_order}) by the spatial eigenmodes $\Xi_k(\xi^{(m)})$, interchanging the order of summation, integrating from $\xi^{(m)}=0$ to $1$ and applying orthonormality of the eigenmodes, yields
\begin{multline}\label{first_order_full_eq}
Q_k^{(m)}\beta^{(m)4}_k+ \frac{\partial^2 Q_k^{(m)}}{\partial T_0^2}=\\
-\bptwo B_k \int_0^1 \frac{\partial V^{(m)}}{\partial T_0}\Xi_k(\xi^{(m)})d\xi^{(m)} 
-\bptwo B_k \frac{\partial \Theta_k^{(m)}}{\partial T_0}
-2 \frac{\partial  ^2 \Theta_k^{(m)}}{\partial T_0 \partial T_1}
\\+
\sum_{n=1}^\infty \left[\bpsev \frac{\partial \Theta_n^{(0)}}{\partial T_0} \mathcal{I}^{k,(m)}_{n,(0)}+\bpsev \frac{\partial \Theta_n^{(1)}}{\partial T_0} \mathcal{I}^{k,(m)}_{n,(1)}+
\bpsix \Theta_n^{(0)} \mathcal{J}^{k,(m)}_{n,(0)}+
\bpsix \Theta_n^{(1)} \mathcal{J}^{k,(m)}_{n,(1)}
\right]\\
\bigg[-\frac{
1}{\pi}\int_0^\pi
\left[
\bpsix \left(\frac{\partial D_{0,SS}}{\partial X}+\frac{\partial V}{\partial X}
\right)
+\bpsev \frac{\partial V}{\partial T_0}
\right]
d\theta \Bigg]
\int_0^1 \cotTm \Xi_k(\xi^{(m)}) d\xi^{(m)}
+ \\  
\sum_{l=1}^\infty 
\bigg[\frac{2}{\pi} \int_0^\pi
\left[
\bpsix \left(\frac{\partial D_{0,SS}}{\partial X}+\frac{\partial V}{\partial X}
\right)
+\bpsev \frac{\partial V}{\partial T_0}
\right]
\cos(l\theta) d\theta \Bigg]
\int_0^1 \sin(l\thm) \Xi_k(\xi^{(m)}) d\xi^{(m)}.
\end{multline}
Equation (\ref{first_order_full_eq}) governs the temporal evolution of each of the spatial eigenmodes. In order to identify secular terms, \citep[which are solutions for the homogeneous equation, see][Ch. 11]{bender1978advanced} 
equation (\ref{Theta_def}) is substituted into equation (\ref{first_order_full_eq}), yielding 
\begin{multline}\label{temp_ref}
Q_k^{(m)}\beta^{(m)4}_k+ \frac{\partial^2 Q_k^{(m)}}{\partial T_0^2}=
-\bptwo B_k \int_0^1 \frac{\partial V^{(m)}}{\partial T_0}\Xi_k(\xi^{(m)})d\xi^{(m)}\\
- \bptwo B_k \beta_k^{(m)2}
\left[
S_k^{(m)}(T_1) \cosk - C_k^{(m)}(T_1) \sink 
\right]
 \\
-2\beta_k^{(m)2} 
\left[
\frac{\partial S_k^{(m)}(T_1) }{\partial T_1} \cosk - \frac{\partial C_k^{(m)}(T_1)}{\partial T_1}  \sink 
\right] 
\\+
\sum_{n=1}^\infty 
\Bigg\{
\bpsev \beta_n^{(0)2}
\bigg[
S_n^{(0)}(T_1) \costz - C_n^{(0)}(T_1) \sintz 
\bigg] \mathcal{I}^{k,(m)}_{n,(0)} 
+ \\
\bpsev \beta_n^{(1)2} \bigg[
S_n^{(1)} (T_1)\costo - C_n^{(1)} (T_1)\sinto  
\bigg]  \mathcal{I}^{k,(m)}_{n,(1)}
+ \\
\bpsix \bigg[ C_n^{(0)}(T_1)\costz
+S_n^{(0)}(T_1)\sintz + \bigg] \mathcal{J}^{k,(m)}_{n,(0)}+\\
\bpsix \bigg[ C_n^{(1)}(T_1) \costo
+S_n^{(1)}(T_1)\sinto\bigg] \mathcal{J}^{k,(m)}_{n,(1)}
\Bigg\} + \hat{F}^{(m)}_k.
\end{multline}

The function $\hat{F}^{(m)}_k$ includes the effect of continuous actuation of the airfoil (i.e. shape modification of airfoil camber at rest by $D_A$). These terms cannot be classified as secular (or not secular) for a general function $D_A$, and need to be assessed separately for any specific actuation function. We thus define $\hatS \sinkin$ and $\hatC \coskin$ as secular terms of $\hat{F}^{(m)}_k$, and other terms as $R_k^{(m)}(T_0)$.  The function $\hat{F}^{(m)}_k$ thus reads
\begin{multline}\label{Fhat}
\hat{F}^{(m)}_k=-\bptwo B_k \int_0^1 \frac{\partial V^{(m)}}{\partial T_0}\Xi_k(\xi^{(m)})d\xi^{(m)}
-\bptwo B_k \frac{\partial \Theta_{k,A}^{(m)}}{\partial T_0} 
+\\ 
\sum_{n=1}^\infty 
\Bigg[
\bpsev\beta_n^{(0)2}
\frac{\partial \Theta_{n,A}^{(0)}}{\partial T_0} 
\mathcal{I}^{k,(m)}_{n,(0)} 
+
\bpsev\beta_n^{(1)2}
\frac{\partial \Theta_{n,A}^{(1)}}{\partial T_0} 
\mathcal{I}^{k,(m)}_{n,(1)}
+
\bpsix  \Theta_{n,A}^{(0)} \mathcal{J}^{k,(m)}_{n,(0)}+
\bpsix  \Theta_{n,A}^{(1)} \mathcal{J}^{k,(m)}_{n,(1)}
\Bigg] 
\\+
\bigg[-\frac{
1}{\pi}\int_0^\pi
\left[
\bpsix \left(
\frac{\partial D_{0,SS}}{\partial X}+\frac{\partial V}{\partial X}
\right)
+\bpsev \frac{\partial V}{\partial T_0}
\right]
d\theta \Bigg]
\int_0^1 \cotTm \Xi_k(\xi^{(m)}) d\xi^{(m)}
+ \\ 
\sum_{l=1}^\infty 
\bigg[\frac{2}{\pi} \int_0^\pi
\left[
\bpsix \left(
\frac{\partial D_{0,SS}}{\partial X}+\frac{\partial V}{\partial X}
\right)
+\bpsev \frac{\partial V}{\partial T_0}
\right]
\cos(l\theta) d\theta \Bigg]
\int_0^1 \sin(l\thm) \Xi_k(\xi^{(m)}) d\xi^{(m)}\\
=R_k^{(m)}+\hatS \sink + \hatC \cosk.
\end{multline}

Substituting (\ref{Fhat}), equation (\ref{temp_ref}) thus takes the form of
\begin{multline}\label{secular_terms_iden}
\frac{\partial^2 Q_k^{(m)}}{\partial T_0^2}
+\beta^{(m)4}_k Q_k^{(m)}=
R_k^{(m)} +\\
\left[
\bptwo B_k\beta_k^{(m)2} C_k^{(m)} (T_1) 
+2\beta_k^{(m)2} \frac{\partial C_k^{(m)} (T_1) }{\partial T_1} 
+\hatS \right] \sink-  \\
\left[
\bptwo B_k\beta_k^{(m)2} S_k^{(m)} (T_1)
+2\beta_k^{(m)2} \frac{\partial S_k^{(m)}(T_1)}{\partial T_1} 
+\hatC \right]\cosk 
+\\
\sum_{n=1}^\infty 
\Bigg[
\bigg[\bpsix S_n^{(0)} (T_1) \mathcal{J}^{k,(m)}_{n,(0)}
 -\bpsev \beta_n^{(0)2} C_n^{(0)}(T_1) \mathcal{I}^{k,(m)}_{n,(0)}
\bigg] \sintz
+ \\
\bigg[ 
\bpsev \beta_n^{(0)2} S_n^{(0)}(T_1) \mathcal{I}^{k,(m)}_{n,(0)} +\bpsix C_n^{(0)}(T_1) \mathcal{J}^{k,(m)}_{n,(0)}
\bigg]  \costz
+ \\
\Bigg[\bpsix S_n^{(1)}(T_1) \mathcal{J}^{k,(m)}_{n,(1)} - \bpsev  \beta_n^{(1)2} C_n^{(1)}(T_1) \mathcal{I}^{k,(m)}_{n,(1)} 
\bigg]\sinto+\\
\bigg[\bpsev \beta_n^{(1)2} S_n^{(1)}(T_1) \mathcal{I}^{k,(m)}_{n,(1)} +\bpsix C_n^{(1)}(T_1) \mathcal{J}^{k,(m)}_{n,(1)}
\bigg]\costo
\Bigg],
\end{multline}
allowing to clearly identify the secular terms of the temporal equation for each of the spatial eigenmodes. 

\subsection{Compatibility equations}
From equation (\ref{secular_terms_iden}) we can obtain compatibility equations and calculate the leading-order temporal behaviour of the orthogonal spatial modes. Depending on the value of $X_c$, the interaction between the front (segment $(0)$) and rear (segment $(1)$) parts of the wing can be separated to 3 distinct cases, using the relation (\ref{freq_relation}): (I) No identical natural oscillation frequencies of the two segments, i.e. $ \beta_k^{(0)} \ne \beta_l^{(1)}\, \forall\, l,k $. (II)  A single identical natural oscillation frequency, i.e. $ \beta_k^{(0)} = \beta_l^{(1)}$, for a single combination of $l\ne k$ (this occurs for infinite number of discrete values of $X_c= {\lambda_k}/{(\lambda_k+\lambda_l)}$, for positive roots $\lambda$ of (\ref{trans_eq})).
 (III) All frequencies coincide, i.e. $ \beta_k^{(0)} = \beta_k^{(1)}\, \forall\, k$ (this occurs only for $X_c=0.5$).

For all cases, the compatibility equation can be obtained by substituting the terms in (\ref{secular_terms_iden}) into a system of first-order ordinary differential equations for $S_n^{(m)}$ and $C_n^{(m)}$. This yields a compatibility condition in the form
\begin{equation}\label{compatabily_matrix_corm}
\frac{\partial\boldsymbol{F}(T_1)}{\partial T_1}=
[\boldsymbol{A}] \boldsymbol{F}(T_1) + \boldsymbol{b}.
\end{equation}
Solutions of (\ref{compatabily_matrix_corm}) are given by
\begin{equation}\label{ODEs_solution}
    \boldsymbol{F}=[\hat{\boldsymbol{A}}(T_1)] 
    \left( [\hat{\boldsymbol{A}}(0)]^{-1} \boldsymbol{F}(0)
    +\int_0^{T_1}[\hat{\boldsymbol{A}}(\tau)]^{-1} \boldsymbol{b} d\tau \right)
\end{equation}
where $[\hat{\boldsymbol{A}}(T_1)]$ is a fundamental matrix solution, given by
\begin{equation}
    [\hat{\boldsymbol{A}}(T_1)]=
    \begin{bmatrix}
    \vdots & \vdots & & \vdots \\
    \boldsymbol{v}_1 e^{\sigma_1 T_1},&
    \boldsymbol{v}_2 e^{\sigma_2 T_1},&
    \cdots, &
    \boldsymbol{v}_n e^{\sigma_n T_1}\\
    \vdots & \vdots & & \vdots 
    \end{bmatrix}
\end{equation}
and $\sigma_i$ and $\boldsymbol{v}_i$ ($i\in\{1,2,...\}$) are eigenvalues and associated eigenvectors of $[\boldsymbol{A}]$, respectively.

Therefore, the temporal eigenmodes involve terms such as $e^{\sigma_i T_1}\cos\left({\beta^{(m)2}_n T_0}\right)$. Thus, stability requires $\Real{[\sigma_i]}<0$ and slow-scale modulation of the elastic-inertial oscillations due to aeroelastic dynamics are related to $\Imag{[\sigma_i]}$. Solutions of the compatibility equation for cases I, II and III are presented below.

\subsubsection{Case (I): $X_c\neq {\lambda_k} /(\lambda_k+\lambda_l)$} 

\begin{subequations}\label{no_fr}
In the absence of identical oscillation frequencies of the front and rear parts of the airfoil, compatibility equations may be obtained separately for each segment. The matrix $[\boldsymbol{A}]$ for segment $(m)$ is this case is thus
\begin{equation}
[\boldsymbol{A}]=
\frac{1}{2}
\begin{bmatrix}
\bpsev \mathcal{I}^{k,(m)}_{k,(m)} -\bptwo B_k &
-{\bpsix \mathcal{J}^{k,(m)}_{k,(m)}}/{\beta_k^{(m)2}} \\
{\bpsix \mathcal{J}^{k,(m)}_{k,(m)}}/{\beta_k^{(m)2}} &
\bpsev \mathcal{I}^{k,(m)}_{k,(m)} -\bptwo B_k
\end{bmatrix}
\end{equation}
and the vectors $\boldsymbol{F}$ and $\boldsymbol{b}$ are  given by
\begin{equation}
\boldsymbol{F}=
\begin{bmatrix}
C_k^{(m)} \\ S_k^{(m)}
\end{bmatrix},
\qquad
\boldsymbol{b}=
-\frac{1}{2\beta_k^{(m)2}}
\begin{bmatrix}
\hatS \\ \hatC
\end{bmatrix}.
\end{equation}
\end{subequations}

The eigenvalues of $[\boldsymbol{A}]$ are thus
\begin{equation}\label{eigenvals}
    \frac{\bpsev \mathcal{I}^{k,(m)}_{k,(m)} -\bptwo B_k}{2}
    \pm i\frac{\bpsix \mathcal{J}^{k,(m)}_{k,(m)}}{2\beta_k^{(m)2}} 
\end{equation}
with associated eigenvectors $\begin{bmatrix} \pm i, & 1 \end{bmatrix}^t$. The real part of the eigenvalues (\ref{eigenvals}) represents the growth or decay of a given mode and  the imaginary part represents the additional slow oscillation due to aeroelastic effects. 

\begin{figure}
\centering 
\includegraphics[width=0.8\textwidth]{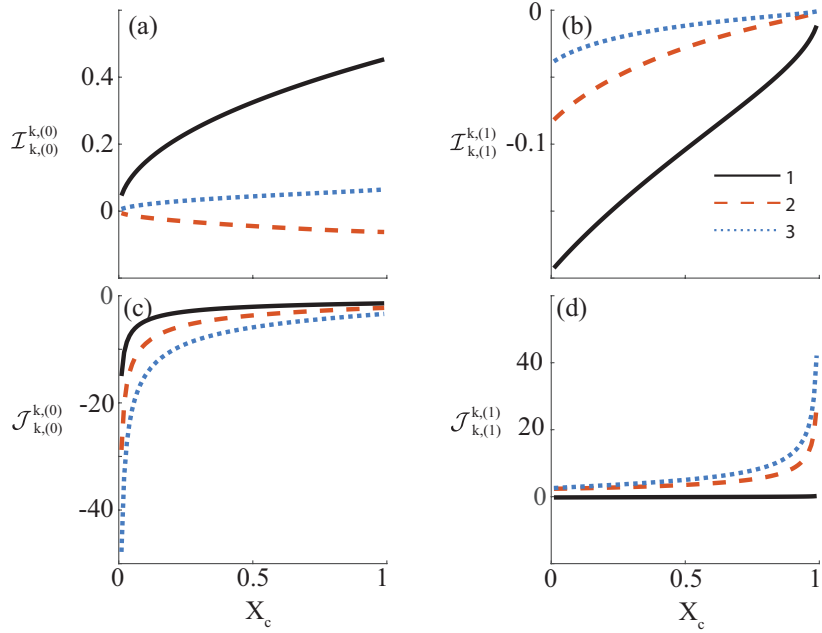}
\caption{Values of $\mathcal{I}^{k,(0)}_{k,(0)}$ (panel (a)), $\mathcal{I}^{k,(1)}_{k,(1)}$ (panel (b)), $\mathcal{J}^{k,(0)}_{k,(0)}$ (panel (c)) and $\mathcal{J}^{k,(1)}_{k,(1)}$ (panel (d)), as a function of the clamping location $X_c$, for $k\in\left\{1,2,3\right\}$ (see  (\ref{operator_defenition})).
Values of $\mathcal{I}^{k,(m)}_{k,(m)}$ represent the self-stability of each mode. It can be seen that the downstream part ($m=1$) is stable, as expected, while the upstream part ($m=0$) is {stable} for even modes, and unstable for odd modes.
Values of $\mathcal{J}^{k,(m)}_{k,(m)}$ represent the corresponding modulation frequencies.}\label{IJ_values}
\end{figure}

The homogeneous part of the solution for $\boldsymbol{F}$, with substituted initial values (\ref{IVS_C&S}), is 
\begin{multline}\label{no_identical_homogeneous_sol}
    \begin{bmatrix}
C_k^{(m)} \\ S_k^{(m)}
\end{bmatrix} =
\begin{bmatrix}
C_k^{(m)}(0) \cos \left(\frac{\Pi_6 \mathcal{J}^{k,(m)}_{k,(m)}}{2\beta_k^{(m)2}} T\right)
-S_k^{(m)}(0) \sin \left(\frac{\Pi_6 \mathcal{J}^{k,(m)}_{k,(m)}}{2\beta_k^{(m)2}} T\right)
\\
S_k^{(m)}(0) \cos\left(\frac{\Pi_6 \mathcal{J}^{k,(m)}_{k,(m)}}{2\beta_k^{(m)2}} T\right)
+ C_k^{(m)}(0) \sin\left(\frac{\Pi_6 \mathcal{J}^{k,(m)}_{k,(m)}}{2\beta_k^{(m)2}} T\right)
\end{bmatrix}
e^{\frac{\Pi_7 \mathcal{I}^{k,(m)}_{k,(m)} -\Pi_2 B_k}{2}T}.
\end{multline}
The values of $\mathcal{I}^{k,(m)}_{k,(m)}$ (representing growth or decay of instability) and $\mathcal{J}^{k,(m)}_{k,(m)}$ (which represents slow modulation frequencies), are  defined in equation (\ref{operator_defenition}) and are functions of $X_c$ only. The first $3$ modes are presented in figure \ref{IJ_values}. The parameter $\mathcal{I}^{k,(0)}_{k,(0)}$, representing the stability of the front region for mode $k$, is positive (increasing instability) for odd modes, and negative for even modes. In contrast, the rear part $\mathcal{I}^{k,(1)}_{k,(1)}$ is inherently stable for all modes, representing aeroelastic damping. Thus, instability emanates from odd modes of the front region of the wing. The parameters $\mathcal{J}^{k,(0)}_{k,(0)}$ and $\mathcal{J}^{k,(1)}_{k,(1)}$ associated with slow aeroelastic frequencies increase, as expected, as the elastic region they represent shorten due to larger values of $X_c$.

\subsubsection{Cases (II) and (III):  $X_c={\lambda_k} /(\lambda_k+\lambda_l)$}
\begin{subequations}\label{one_fr}

In Case (II)  the frequency of mode $k$ of the front segment $\beta_k^{(0)}$ equals the frequency of mode $l$ of the rear segment $\beta_l^{(1)}$. Thus, a single compatibility condition governs mode $k$ of the front segment and mode $l$ of the rear segment, yielding the $[\boldsymbol{A}]$ matrix of
\begin{equation}\label{nkeigen}
[\boldsymbol{A}]=
\frac{1}{2}
\begin{bmatrix}
 \bpsev \mathcal{I}^{k,(0)}_{k,(0)} -\bptwo B_k          &
-{\bpsix \mathcal{J}^{k,(0)}_{k,(0)}}/{\tilde{\beta}^2}&
 \bpsev \mathcal{I}^{k,(0)}_{l,(1)}                      &
-{\bpsix \mathcal{J}^{k,(0)}_{l,(1)}}/{\tilde{\beta}^2}
\\
 {\bpsix \mathcal{J}^{k,(0)}_{k,(0)}}/{\tilde{\beta}^2}&
 \bpsev \mathcal{I}^{k,(0)}_{k,(0)} -\bptwo B_k          &
 {\bpsix \mathcal{J}^{k,(0)}_{l,(1)}}/{\tilde{\beta}^2}& 
 \bpsev \mathcal{I}^{k,(0)}_{l,(1)}
 \\
 \bpsev \mathcal{I}^{l,(1)}_{k,(0)}                      &
-{\bpsix \mathcal{J}^{l,(1)}_{k,(0)}}/{\tilde{\beta}^2}&
 \bpsev \mathcal{I}^{l,(1)}_{l,(1)} -\Pi_2 B_l           &
-{\bpsix \mathcal{J}^{l,(1)}_{l,(1)}}/{\tilde{\beta}^2}
\\
 {\bpsix \mathcal{J}^{l,(1)}_{k,(0)}}/{\tilde{\beta}^2}&
 \bpsev \mathcal{I}^{l,(1)}_{k,(0)}                      &
 {\bpsix \mathcal{J}^{l,(1)}_{l,(1)}}/{\tilde{\beta}^2}&
 \bpsev \mathcal{I}^{l,(1)}_{l,(1)} -\bptwo B_l
\end{bmatrix}
\end{equation}
where $\beta_k^{(0)}=\beta_l^{(1)}=\tilde{\beta}$, and the related $\boldsymbol{F}$ and $\boldsymbol{b}$ vectors are
\begin{equation}
\boldsymbol{F}=
\begin{bmatrix}
C_k^{(0)} \\ S_k^{(0)} \\ C_l^{(1)} \\ S_l^{(1)}
\end{bmatrix},
\qquad
\boldsymbol{b}=
-\frac{1}{2\tilde{\beta}^2}
\begin{bmatrix}
\hat{S}_k^{(0)} \\ \hat{C}_k^{(0)} \\ \hat{S}_l^{(1)} \\ \hat{C}_l^{(1)}
\end{bmatrix}.
\end{equation}
\end{subequations}
Matrix $[\boldsymbol{A}]$ thus involves operator products such as $\mathcal{I}^{k,(0)}_{l,(1)}$, representing interaction between the elastic oscillation mode $k$ of the front segment and the aerodynamic forces due to mode $l$ oscillations of the rear segment.  Unlike the previous case (I), we cannot deduce stability directly from the values of terms such as $\mathcal{I}^{k,(0)}_{l,(1)}$ and require computation of the eigenvalues of $[\boldsymbol{A}]$.



Similarly, case (III) occurring for $X_c=0.5$ represents identical frequencies of the front and the rear parts for all modes, i.e. $\beta_k^{(0)} = \beta_k^{(1)} = \beta_k = 2\lambda_k\, \forall\,k$. In this case, for all modes $k$, matrix $[\boldsymbol{A}]$ is of the form
\begin{subequations}\label{all_fr}
\begin{equation}\label{all_frA}
[\boldsymbol{A}]=
\frac{1}{2}
\begin{bmatrix}
 \bpsev \mathcal{I}^{k,(0)}_{k,(0)} -\bptwo B_k     &
-{\bpsix \mathcal{J}^{k,(0)}_{k,(0)}}/{\beta_k^2} &
 \bpsev \mathcal{I}^{k,(0)}_{k,(1)}                 &
-{\bpsix \mathcal{J}^{k,(0)}_{k,(1)}}/{\beta_k^2}
\\
 {\bpsix \mathcal{J}^{k,(0)}_{k,(0)}}/{\beta_k^2} &
 \bpsev \mathcal{I}^{k,(0)}_{k,(0)} -\bptwo B_k     &
 {\bpsix \mathcal{J}^{k,(0)}_{k,(1)}}/{\beta_k^2} &
 \bpsev \mathcal{I}^{k,(0)}_{k,(1)}
 \\
 \bpsev \mathcal{I}^{k,(1)}_{k,(0)}                 &
-{\bpsix \mathcal{J}^{k,(1)}_{k,(0)}}/{\beta_k^2} &
 \bpsev \mathcal{I}^{k,(1)}_{k,(1)} -\bptwo B_k      &
-{\bpsix \mathcal{J}^{k,(1)}_{k,(1)}}/{\beta_k^2}
\\
 {\bpsix \mathcal{J}^{k,(1)}_{k,(0)}}/{\beta_k^2} &
 \bpsev \mathcal{I}^{k,(1)}_{k,(0)}                 &
 {\bpsix \mathcal{J}^{k,(1)}_{k,(1)}}/{\beta_k^2} &
 \bpsev \mathcal{I}^{k,(1)}_{k,(1)} -\bptwo B_k
\end{bmatrix}
\end{equation}
and the vectors $\boldsymbol{F}$ and $\boldsymbol{b}$ are
\begin{equation}
\boldsymbol{F}=
\begin{bmatrix}
C_k^{(0)} \\ S_k^{(0)} \\ C_k^{(1)} \\ S_k^{(1)}
\end{bmatrix}, \qquad
\boldsymbol{b}=
-\frac{1}{2\beta_k^2}
\begin{bmatrix}
\hat{S}_k^{(0)} \\ \hat{C}_k^{(0)} \\ \hat{S}_k^{(1)} \\ \hat{C}_k^{(1)}
\end{bmatrix}.
\end{equation}
\end{subequations}
In this case matrix $[\boldsymbol{A}]$ involves the operator products such as $\mathcal{I}^{k,(0)}_{k,(1)}$, representing interaction between the elastic oscillation mode $k$ of the front segment and the aerodynamic forces due to the same mode $k$ of the rear segment.

\subsection{Stability condition}
The compatibility equations derived in \S 4.4 yield the growth or decay of the various modes of the rear and front segments, or interaction between the modes of the segments,  for arbitrary values of $X_c$ via computation of the eigenvalues of $[\boldsymbol{A}]$.  Furthermore, in $[\boldsymbol{A}]$ the damping terms, denoted by $B$, appear only on the diagonal of $[\boldsymbol{A}]$. Thus, we can compute eigenvalues $\sigma$ from the characteristic polynomial $p(\sigma)=|[A]-\sigma I|$ for $B=0$, and then directly obtain the value of ${\Pi_2 B}/{2}$ required for stability.

We denote hereafter $\tilde\sigma$ as the eigenvalue, calculated for $B=0$, with the maximal real part. This eigenvalue corresponds to the mode, or interaction between two modes, with the maximal growth rate which eventually dominates the dynamics of the configuration. Thus, the stability condition is
\begin{equation}
    \frac{\Pi_2 B}{2}=\frac{c^2 r_d}{2\sqrt{s^*\mu_s^*}}>\Real{[\tilde\sigma]}.
\end{equation}

For case (I) $X_c\neq{\lambda_k}/(\lambda_k+\lambda_l)$ (using (\ref{ratios}) ) the dimensional stability condition for spatial mode $k$ and segment $(m)$ is 
\begin{equation}
    2\rho_\infty u_\infty \mathcal{I}^{k,(m)}_{k,(m)} -r^*_{d,k} <0
\end{equation}
where $r^*_{d,k}$ is the dimensional modal damping coefficient for spatial mode $k$. The dimensional aeroelastic modulation frequencies are
\begin{equation}
    \frac{\rho_\infty u_\infty^2 c }
    {\beta_k^{(m)2} \sqrt{\mu^*_s  s^*} } \mathcal{J}^{k,(m)}_{k,(m)}
\end{equation}
and the value of $\tilde\sigma$ is immediately obtained from (\ref{eigenvals}) as $\tilde\sigma=\Pi_7$ $\mathcal{I}^{k,(m)}_{k,(m)}$$/2$. From figure \ref{IJ_values}, the stability condition for the whole system is
\begin{equation} 
    \textcolor{black}{u_\infty < \frac{r^*_{d,k}}{2\rho_\infty \mathcal{I}^{1,(1)}_{1,(1)} }}.
\end{equation}
For cases (II) and (III)  $X_c={\lambda_k}/(\lambda_k+\lambda_l)$, $\tilde\sigma$ needs to be computed from from (\ref{nkeigen}) or (\ref{all_frA}).

\begin{figure}
\centering 
\includegraphics[width=0.8\textwidth]{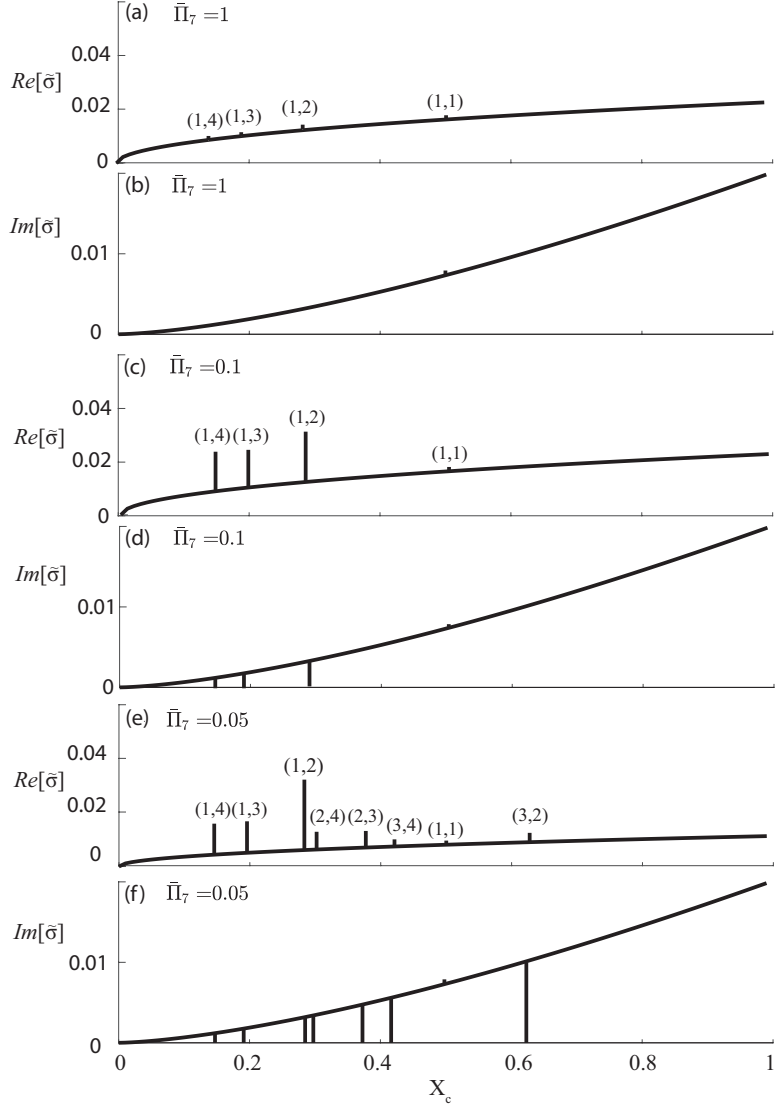}
\caption{Real and imaginary parts of $\tilde\sigma$ vs. $X_c$ for $\bar \Pi_7=1$ (panels (a),(b)), $\bar\Pi_7=0.1$  (panels (c),(d)) and $\bar\Pi_7=0.05$ (panels (e),(f)). In all cases $\bar\Pi_6=1$, taking 4 modes in account. Panels ((a),(c),(e)) present the growth rate of the dominant mode and panels ((b),(d),(f)) present the modulation frequency associated with the dominate mode. Dominant modes associated with interaction between the front and rear modes  are presented by vertical lines due to sudden change in $\tilde\sigma$ at $X_c={\lambda_k}/(\lambda_k+\lambda_l)$. Numbers $(n,k)$ above vertical lines mark the front $n$ and rear $k$ modes interaction which yield the dominant instability growth rate. }
\label{stability_graph}
\end{figure}

Figure \ref{stability_graph} presents the real and imaginary parts of $\tilde\sigma$ vs. $X_c$ for $\Pi_3=1$ and $\bar\Pi_6=1$ for various values of $\bar\Pi_7$, taking 4 modes in account. ($\bar\Pi_7=1$ in panels (a,b), $\bar\Pi_7=0.1$ in panels (c,d) and $\bar\Pi_7=0.05$ in panels (e,f)). Panels (a,c,e) present the growth rate of the dominant mode while panels (b,d,f) present the modulation frequency associated with the dominate mode. Without interaction between the modes, the dominant mechanism of aeroelastic instability in all presented cases emanates from the first mode of the front part of the airfoil. The interaction between the front and rear modes for $X_c={\lambda_k}/(\lambda_k+\lambda_l)$ is presented by vertical lines corresponding to the sudden change in $\tilde\sigma$ at the locations of $X_c$ in which the rear and front modes have identical natural frequencies. Above the vertical lines, the numbers $(n,k)$ mark the front $n$ and rear $k$ modes interaction which yield the dominant instability growth rate. 
For $\bar\Pi_7=1$ the interaction dynamics yield only minor effect on the stability condition (see panel (a)). However, as $\bar\Pi_7$ decreases in comparison to $\bar\Pi_6$, the effect of interaction between modes becomes significant and must be considered when analyzing the stability of such configurations (see panels (c,e)). In addition, the effect of interaction between the front and rear segments increases as $X_c$ decreases and the growth rate of the first front mode decreases. In contrast with modulation frequencies associated with the front segment first mode, instability associated with interaction between modes does not modulate the fast-time oscillation frequency (see panels (d,f)).

\subsection{Dynamic response}
\subsubsection{Initial conditions}
We present the transient response of an elastic airfoil, clamped at $X_c\neq {\lambda_k} /(\lambda_k+\lambda_n)$, to initial conditions which differ from the steady-state solution. Based on \S 4.4, the vector $\boldsymbol{b}$ of the evolution equation as
\begin{equation}\label{no_act_b}
\boldsymbol{b}=
-\frac{1}{2\beta_k^{(m)2}}
\begin{bmatrix}
\hatS \\ \hatC
\end{bmatrix}=
\begin{bmatrix}
0\\0
\end{bmatrix}.
\end{equation}

Substituting (\ref{no_act_b}) and the compatibility condition (\ref{no_identical_homogeneous_sol}) into the leading-order solution (\ref{LE_sol}), yields the multi-scale leading order solution as $D_0^{(m)}$ as
\begin{subequations}\label{ssss}
\begin{multline}
    D_0^{(m)}=D_{0,SS}^{(m)}+
    \sum_{n=1}^\infty 
    \Bigg\{ 
    \Xi_n(\xi^{(m)})
    exp\left[{\frac{\left(\Pi_7 \mathcal{I}^{n,(m)}_{n,(m)} -\Pi_2 B_n\right)T}{2}}\right] \times \\
    \bigg[
    \left(
    C_n^{(m)}(0) \cos\left(\frac{\Pi_6 \mathcal{J}^{n,(m)}_{n,(m)}T}{2\beta_n^{(m)2}} \right)
    -S_n^{(m)}(0) \sin\left(\frac{\Pi_6 \mathcal{J}^{n,(m)}_{n,(m)}T}{2\beta_n^{(m)2}} \right)
    \right)
    \cos \left(\beta_n^{(m)2}T\right) 
    \\+
    \left(
    S_n^{(m)}(0) \cos\left(\frac{\Pi_6 \mathcal{J}^{n,(m)}_{n,(m)}T}{2\beta_n^{(m)2}} \right)
    +C_n^{(m)}(0)\sin\left(\frac{\Pi_6 \mathcal{J}^{n,(m)}_{n,(m)}T}{2\beta_n^{(m)2}} \right)
    \right)
    \sin \left(\beta_n^{(m)2}T\right)
    \bigg]
    \Bigg\},
\end{multline}
where
\begin{equation}
     C_n^{(m)}(0)=\int_0^1 \left[F_1(\xi^{(m)})-D_{0,SS}(\xi^{(m)})\right]\Xi_n(\xi^{(m)})d\xi^{(m)}
\end{equation}
and
\begin{equation}
    S_n^{(m)}(0)=\frac{1}{\beta^{(m)2}_n}\int_0^1 F_2(\xi^{(m)})\Xi_n(\xi^{(m)})d\xi^{(m)}.
\end{equation}
\end{subequations}

\begin{figure}
\centering 
\includegraphics[width=1\textwidth]{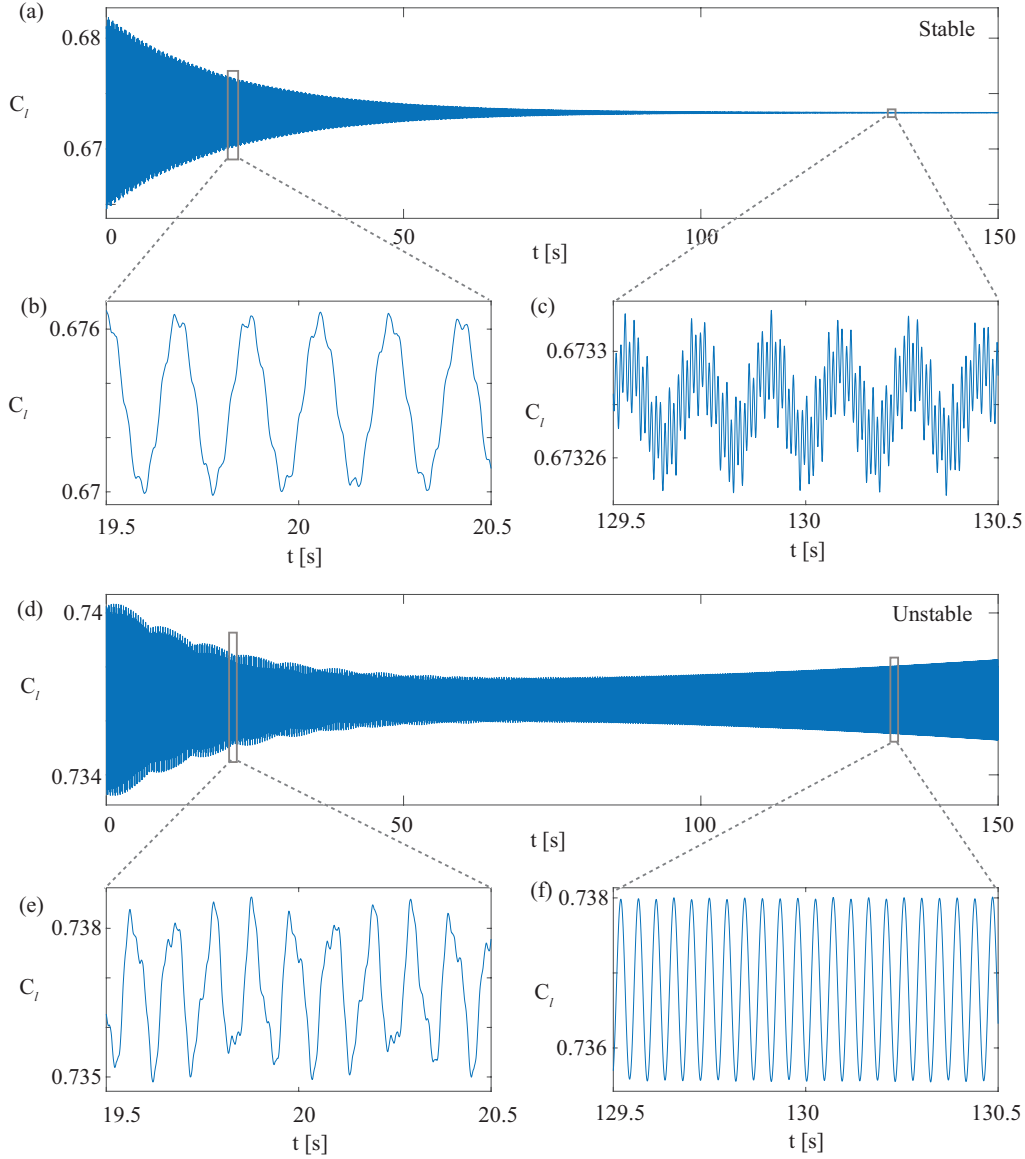}
\caption{Lift coefficient $C_l$ vs. time computed for an elastic unactuated NACA4405 airfoil clamped at $x_c=0.2c$ (stable configuration, panels (a-c)) and $x_c=0.4c$ (unstable configurations, panels (d-f)). Relevant parameters are  $E_{eq}=10^9[Pa]$, $s=E_{eq}(0.05c)^3/12[Pa\times m^2]$, $\mu_{s,eq}=270 [{kg}/m]$, $c=1[m]$, $\rho_\infty= 1.2 [{kg}/{m^3}]$,  $u_\infty=30[{m}/{s}]$, and $\alpha=3^\circ$, yielding dimensionless numbers $\bptwo=0.01$, $\Pi_4=1$, $\Pi_5=0.76$, $\bpsix=0.21$, and $\bpsev=0.04$. Initial conditions are $D(T=0,X)=0$ and $\partial D/\partial T (T=0,X)=0$.}
\label{no_inlet_graphs}
\end{figure}

Figure \ref{no_inlet_graphs} presents the lift coefficient (defined as $C_l=2l/\rho_\infty
u_\infty^2c$, where $l$ is lift-per-unit-span)  vs. time computed from (\ref{ssss}) for a NACA4405 elastic airfoil, with dimensional average Young's modulus $E_{eq}=10^9[Pa]$, rigidiy $s=E_{eq}(0.05c)^3/12[Pa\times m^2]$, solid mass per-unit-lengh $\mu_{s,eq}=270 [{kg}/m]$, chord length $c=1[m]$, air density $\rho_\infty= 1.2 [{kg}/{m^3}]$, air speed  $u_\infty=30[{m}/{s}]$, and angle-of-attack $\alpha=3^\circ$. The above dimensional parameters yield the dimensionless ratios $\Pi_2=0.01$, $\Pi_4=1$, $\Pi_5=0.76$, $\Pi_6=0.21$, and $\Pi_7=0.04$. The airfoil is initially at rest, $D(T=0,X)=0$ and $\partial D/\partial T (T=0,X)=0$, and deformations occur due to aerodynamic loads on the profile. Panel (a) present a stable configuration with $x_c=0.2c$, and panels (b) and (c) present closeups on early and late stages. 
Panel (d) presents an unstable configuration with $X_c=0.4c$, and similarly panels (e) and (f) present closeups. For both configurations early times ($t\leq50[s]$) involves gradual decay of the initial excitation due to the effect of the external loads. Closeups (b) and (e) show multiple overlapping modes during the initial gradual decay. For the stable configurations, all modes continue to decay and appear in late times, as evident in panel (c), while the in unstable configuration a single mode grows and dominate the dynamics, as evident in panel (f).



\subsubsection{Oscillatory actuation}
We present the effect of distributed actuation of the form
\begin{equation}\label{DA_a}
    D_A(X,T_0)=-\frac{1}{2} \left(X-X_c\right)^2 \sin\left(\kappa T_0\right),
\end{equation}
which may represent uniform distributed shape-morphing actuation methods.



Substituting (\ref{DA_a}) into (\ref{V_def}), the corresponding homogenization function $V^{(m)}$ is
\begin{equation}
    V^{(m)}=\frac{\Pi_1}{2} \left(X-X_c\right)^2 \sin\left(\kappa T_0\right)
    =\frac{\Pi_1}{2} \left(m-X_c\right)^2 \xi^{(m)2} \sin\left(\kappa T_0\right),
\end{equation}
and substituting into (\ref{Theta_a}) yields 
\begin{equation}
    \Theta_{n,A}^{(m)}= 
    \frac{\Pi_1  \kappa^2 \left(m-X_c\right)^2}{2 \beta_n^{(m)2}}
    \left[
    \sint \circledast \sin\left(\kappa T_0\right)
    \right]
    \int_0^1 \Xi_n(\xi^{(m)})\xi^{(m)2}d\xi^{(m)}.
\end{equation}
The convolution product for a specific mode $k$ and segment $(m)$ depends on the actuation frequency. For actuation frequency which is not the natural mode frequency, i.e. $\kappa \ne \beta_n^{(m)2}$, we obtain
\begin{equation}\label{notfr}
    \sint \circledast \sin\left(\kappa T_0\right)=
     \frac{\kappa \sint -\beta_n^{(m)2} \sin\left(\kappa T_0\right)}{\kappa^2 -\beta_n^{(m)4}}.
\end{equation} 
However, as $\kappa \rightarrow \beta_n^{(m)2}$, (\ref{notfr}) becomes singular, and the convolution yields the limit
\begin{equation}
    \sint \circledast \sint=
    \frac{1}{2}\left[ \frac{1}{\beta_n^{(m)2}}\sint -T_0 \cost\right]
\end{equation} 
which, as expected, modifies the compatibility equation (appearing within $\boldsymbol{b}$).  Thus, for the non-resonance input frequency case $\kappa \ne \beta_n^{(m)2}$, vector $\boldsymbol{b}$ is
\begin{equation}
    \boldsymbol{b}=
-\frac{1}{2\beta_k^{(m)2}}
\begin{bmatrix}
{\bpsix \mathcal{J}^{k,(m)}_{k,(m)}}/{\beta_k^{(m)2}}
\\
\bpsev \beta_k^{(m)2} \mathcal{I}^{k,(m)}_{k,(m)} +\bptwo B_k 
\end{bmatrix}
\frac{\Pi_1 \kappa^3 (m-X_c)^2}{2(\beta_k^{(m)4}-\kappa^2)}\int_0^1 \Xi_k (\xi^{(m)})\xi^{(m)2}d\xi^{(m)},
\end{equation}
and for the resonance input frequency $\kappa = \beta_n^{(m)2}$, vector $\boldsymbol{b}$ is
\begin{multline}
      \boldsymbol{b}=
\begin{bmatrix}
(\bptwo B_k-\bpsev \mathcal{I}^{k,(m)}_{k,(m)}\beta_k^{(m)2})\beta_k^{(m)2}\\
\bpsix  \mathcal{J}^{k,(m)}_{k,(m)}
\end{bmatrix}T_0
\frac{\Pi_1 (m-X_c)^2}{4}\int_0^1 \Xi_k (\xi^{(m)})\xi^{(m)2}d\xi^{(m)}+\boldsymbol{c},
\end{multline}
where $\boldsymbol{c}$ is vector containing only constants.

The compatibility condition (\ref{compatabily_matrix_corm}) for nonzero $\boldsymbol b$ is  given by (\ref{ODEs_solution})
where the system's fundamental matrix $\left[\boldsymbol{\hat{A}}(T_1)\right]$ is:
\begin{equation}
    \left[\boldsymbol{\hat{A}}(T_1)\right]=
    \begin{bmatrix}
    -\sin\left(\frac{\bpsix \mathcal{J}^{n,(m)}_{n,(m)}}{2} T_1\right) &
    \cos\left(\frac{\bpsix \mathcal{J}^{n,(m)}_{n,(m)}}{2} T_1\right) \\
    \cos\left(\frac{\bpsix \mathcal{J}^{n,(m)}_{n,(m)}}{2} T_1\right) &
    \sin\left(\frac{\bpsix \mathcal{J}^{n,(m)}_{n,(m)}}{2} T_1\right)
    \end{bmatrix}
    e^{\frac{\bpsev \mathcal{I}^{n,(m)}_{n,(m)} -\bptwo B_n}{2}T_1} 
\end{equation}

\begin{figure}
\centering 
\includegraphics[width=1\textwidth]{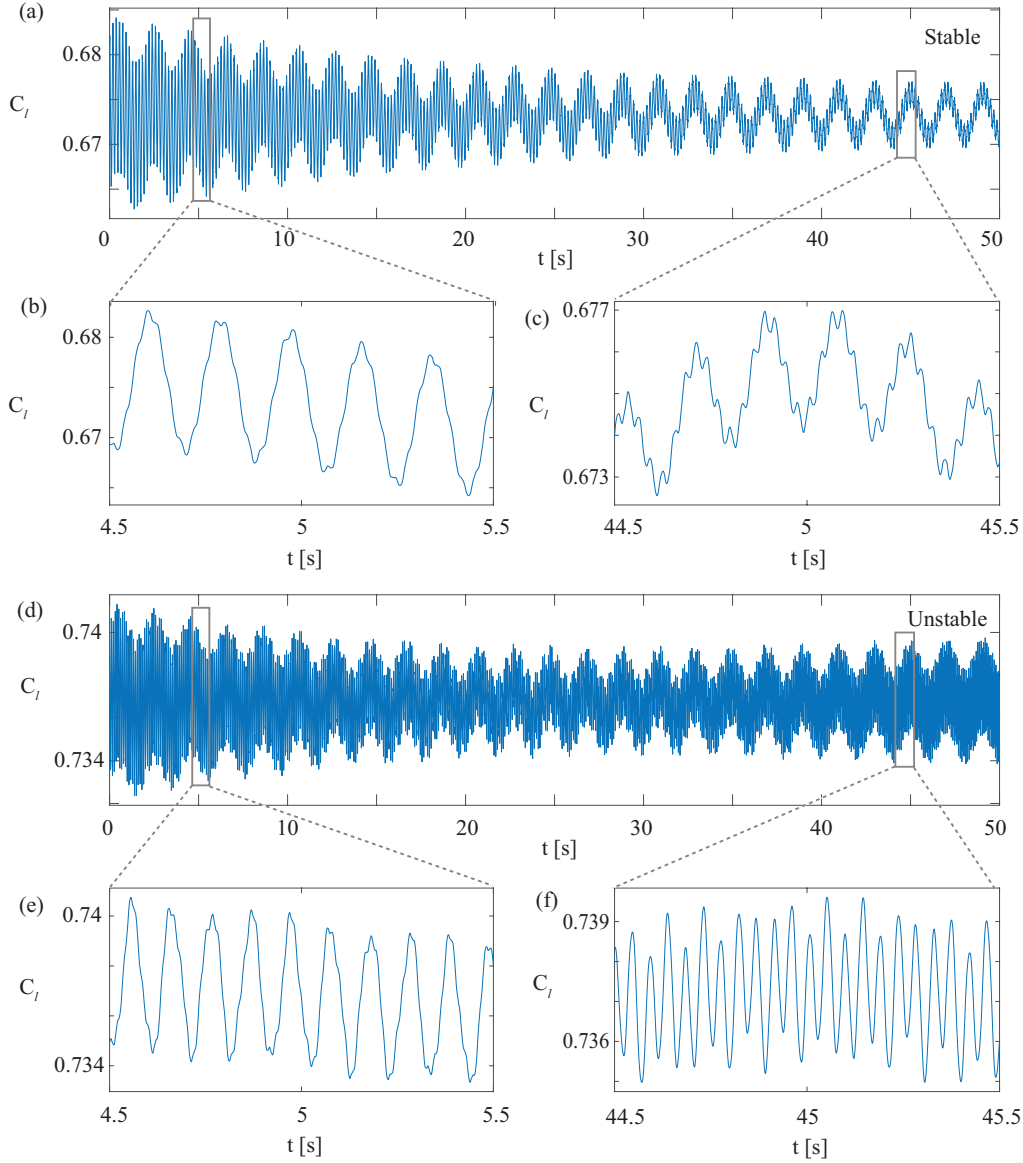}
\caption{Lift coefficient $C_L$ vs. time for oscillatory actuation $D_A(X,T)=-\left(X-X_c\right)^2 \sin\left(\kappa T_0\right)/2$ computed for an elastic unactuated NACA4405 airfoil clamped at $x_c=0.2c$ (stable configuration, panels (a-c)) and $x_c=0.4c$ (unstable configurations, panels (d-f)). Actuation frequency is $\kappa=0.5$ and actuation amplitude is $\Pi_1=0.92$. All other parameters and initial conditions are identical to figure \ref{no_inlet_graphs}.}
\label{Sine_inlet_graphs}
\end{figure}

Figure \ref{Sine_inlet_graphs} presents oscillations of lift coefficient (defined as $C_l=2l/\rho_\infty
u_\infty^2$, where $l$ is lift-per-unit-span)  vs. time for a NACA4405 elastic airfoil actuated at the form (\ref{DA_a}) with nondimensional frequency of $\kappa=0.5$ ($\kappa\neq\beta_n^{(m)2}$) and amplitude $\Pi_1=d_a^*/d^*=0.92$.
Initial conditions, as well as all dimensional parameters, are identical to those presented in \S 4.6.1. In panels (a-c) $x_c=0.2c$ and in panels (e-f) $x_c=0.4c$.  Panel (a), and closeups (b) and (c), present a stable configuration. Initially, the forced actuation is negligible compared with the modes excited by the aerodynamic forcing. This is reversed in late times where all natural oscillations modes decay leaving $D_a$ as the dominant oscillation. Panel (d) and closeups (e) and (f) present an unstable configuration (due to the location $x_c=0.4c$ and the first mode of the front segment). Initial decay of multiple modes is observed, similarly to figure \ref{no_inlet_graphs}, transitioning to gradual growth of a single dominant mode coupled with the actuation forcing for late times.

\section{Concluding remarks}
This work presented analysis and numerical calculations of a shape-morphing soft two-dimensional  airfoil in potential flow. The airfoil was modelled as two cantilevered elastic sheets connected to a rigid support at an arbitrary location amid chord. Steady-state and transient solutions are presented, based on regular and matched asymptotics, respectively. Stability conditions, and initial dynamics of stable and unstable configurations, are obtained from the compatibility equations of the different spatial modes. The maximal stable speed is presented as a function of elastic damping, fluid density and location of clamping. Focus is given to the interaction between the front and rear segments, which is shown to be a dominant instability mechanism for a set of discrete locations of clamping. 
The presented results lay a theoretical foundation for the realization of shape-morphing soft airfoils.

The most limiting simplifying assumption used in the current study was neglecting vortex shedding effects. While this assumption is commonly used, and significantly simplifies the  analysis of such configurations, future research is required to assess the effect of vortex shedding on such soft airfoils. In addition, in the current study the clamping location, $X_c$, is taken as constant for a given configuration. However, the main importance of $X_c$ in the analysis is setting the natural frequencies of the front and rear segments. Significant interaction effects occur for identical frequencies of one of the spatial modes of the front segment and one of the modes of the rear segment. However, actuation of soft airfoils is expected to change the properties of the front and rear segments, and thus change the natural frequencies during actuation. Thus, instability due to intersection of front and rear natural frequencies may occur during actuation.

\acknowledgments{We thank Dr. Sonya Tiomkin and Prof. Daniella Raveh for helpful discussions.}


\end{document}